\documentclass[11pt,a4paper]{article}
\pdfoutput=1
\usepackage{graphicx}
\usepackage{jheppub}
\usepackage{bm}
\usepackage{multirow}
\newcommand{\beq}{\begin{equation}}
\newcommand{\eeq}{\end{equation}}
\newcommand{\bea}{\begin{eqnarray}}
\newcommand{\eea}{\end{eqnarray}}
\newcommand{\bpmat}{\begin{pmatrix}}
\newcommand{\epmat}{\end{pmatrix}}

\title{Standard Coupling Unification in SO(10), Hybrid Seesaw
   Neutrino Mass and Leptogenesis, Dark Matter, and Proton
  Lifetime Predictions}

\author{M. K. Parida$^{1,*}$, Bidyut Prava Nayak$^{1}$, Rajesh
  Satpathy$^{1}$, Ram Lal
  Awasthi$^{2,***}$ }
\affiliation{${}^1$Centre of Excellence in Theoretical and Mathematical Sciences\\
Siksha `O' Anusandhan University, Khandagiri Square\\
 Bhubaneswar 751030, India\\

${}^2$Indian Institute of Science Education and Research\\
Knowledge City, Sector 81, SAS Nagar, Manauli 140306, India}
\emailAdd{${}^*$minaparida@soauniversity.ac.in}
\emailAdd{${}^{***}$awasthi.r6@gmail.com}

\abstract{ We discuss gauge coupling unification of
   $SU(3)_C\times SU(2)_L\times U(1)_Y$  descending directly
  from non-supersymmetric SO(10) while providing solutions to the
  three outstanding problems of the standard model: neutrino masses,
  dark matter, and the baryon  asymmetry of the universe. Conservation
  of matter parity as gauged discrete symmetry for the stability and  identification of dark matter in the model
  calls for high-scale spontaneous symmetry breaking through ${126}_H$
  Higgs representation. This naturally leads to the hybrid seesaw
  formula for neutrino masses  mediated by heavy scalar triplet and
  right-handed neutrinos. Being quadratic in the Majorana coupling, the seesaw formula predicts
two distinct patterns of right-handed neutrino masses, one hierarchical and
  another not so hierarchical (or compact), 
when  fitted with  the neutrino oscillation data.  Predictions of the baryon asymmetry via leptogenesis are investigated through the decays of both the patterns of RH$\nu$ masses. A complete flavor analysis has been carried out to compute CP-asymmetries including washouts and solutions to  Boltzmann equations have been utilised  to predict the baryon asymmetry. The additional contribution to vertex correction mediated by the heavy left-handed triplet scalar is  noted to 
contribute as dominantly as other Feynman diagrams. We have found
successful predictions of the baryon asymmetry for both the patterns
of right-handed neutrino masses. 
The
 $SU(2)_L$ triplet fermionic dark matter at the TeV scale carrying even matter
  parity is naturally embedded into the non-standard fermionic
  representation ${45}_F$ of SO(10). In addition to the triplet scalar and
  the triplet fermion, the model needs a
  nonstandard color octet fermion of mass $\sim 5 \times 10^{7}$ GeV  to achieve
  precision gauge coupling unification at the GUT mass scale
  $M_U^0=10^{15.56}$ GeV. Threshold corrections due to
  superheavy components of  ${126}_H$ and other representations are
  estimated and found to be substantial. It is noted that the proton
  life time predicted by the model is accessible to the ongoing and
  planned experiments over a wide range of parameter space.} 

\begin{document}
\maketitle

%%%%%%%%%%%%%%%%%%%%%%%%%%%%%%%%%%%%%%%%%%%%%%%%%%%%%%%%%%%%%%%%%%% 
\section{Introduction}\label{sec.1}
The standard model (SM) of particle interactions based upon the gauge
symmetry $SU(3)_C\times SU(2)_L \times U(1)_Y$ has been tested by
numerous experiments. Also the last piece of evidence in
favour of the SM has been vindicated with the
discovery of the Higgs boson at the CERN Large Hadron Collider \cite{Higgsexpt:2012}. Yet the model fails to
explain
 the three  glaring physical phenomena: neutrino oscillation \cite{nudata},  baryon asymmetry of
the universe (BAU) \cite{BAUexpt,Planck15}, and  dark matter (DM)
\cite{DMexpt}. Although the
electroweak part of the SM provides excellent description of weak
interaction phenomenology manifesting in $V-A$ structure of neutral
and charged currents,  it fails
to answer why parity violation is exhibited by weak interaction alone.
 On the fundamental side, the SM itself can not
explain the disparate values of its gauge couplings.
 The minimal
gauge theory which has the potential to unify the three gauge
couplings \cite{GQW:1974,su5} and explain the origin of parity violation is $SO(10)$ grand
unified theory (GUT) \cite{so10} that contains the Pati-Salam
\cite{Pati-Salam:1974} and left-right gauge theories \cite{rnmjcp:1975} as
its subgroups. However, it is well known that direct breaking of all
non-supersymmetric (non-SUSY) GUTs \cite{so10,su5} to the SM gauge
theory under the assumption of minimal fine tuning
 hypothesis \cite{del Aguilla:1981,rnm-gs:1983} fails to unify the
 gauge couplings of the SM whereas supersymmetric GUTs like $SU(5)$
 \cite{su5} and
 $SO(10)$ \cite{so10}  achieve this objective in a profound
manner. In fact the prediction of coupling unification in the minimal supersymmetric
 standard model (MSSM) \cite{Dimo-Raby-Wil:1981,Mar-gs:1982} evidenced through the CERN-LEP data
 \cite{Amaldi:1991,Langacker:1991,Ellis:1992} led to the belief that
 a SUSY GUT \cite{Fukuyama:2004ps} with its underlying mechanism for solutions
 to the gauge hierarchy problem
 \cite{Witten:1981,Dimo-Georgi:1982,Romesh:1982} could be the
 realistic model for high energy physics.
SUSY GUTs also predict wino or neutralino as popular candidates of
cold dark matter (CDM). Compared to SUSY
$SU(5)$ \cite{Dimo-Raby-Wil:1981}, SUSY $SO(10)$ has a number of advantages. 
Whereas parity violation in SO(10) has its spontaneous breaking
origin, for SU(5) it is explicit and intrinsic. The right-handed 
neutrino (RH$\nu$) as a member of spinorial representation $16$ of SO(10) mediates the well known
canonical seesaw mechanism \cite{type-I,Valle:1980} that accounts for small
neutrino masses evidenced by the neutrino oscillation data. Further
the Dirac neutrino mass matrix that occurs as an important ingredient
of type-I seesaw \cite{type-I,Valle:1980} is predicted in this model due to its underlying
quark-lepton symmetry \cite{Pati-Salam:1974}. In addition, the presence of
the left-handed (LH) triplet scalar, $\Delta_L (1, 3,-1)\subset
{126}_H\subset SO(10)$, naturally leads to the possibility of Type-II
seesaw formula for neutrino masses \cite{Valle:1980,type-II}. Both the heavy RH
neutrinos and the LH triplet scalar have the high potential  
to account for
BAU via leptogenesis
\cite{Fuku-Yana:1986,Pati:2002,RNM:2002,Hambye:2003}. With R-Parity
as its gauged discrete symmetry \cite{Mohapatra:1986,Krauss:1989,SPMartin:1992,Melfo:1998}, the model
also guarantees stability of dark matter.

 Another  attractive aspect of SUSY SO(10) \cite{baburnm:1993} has been its capability to make a reasonably good representation of all fermion masses and mixings
 at the GUT scale \cite{Bertolini:2006,Joshipura:2011}. Such a data set exhibiting $b-\tau$ Yukawa unification and very approximately satisfying Georgi-Jarlskog \cite{GJ:1979} type relation is obtained using RG extrapolated values of the masses and mixings at the electroweak scale following the bottom-up approach  
 \cite{dp:2001}. In particular  $\chi^2$
 estimation has been carried out to examine goodness of fit to all
 fermion masses in SUSY SO(10) \cite{Joshipura:2011}. Other interesting aspects of
 the SUSY GUT such as Yukawa unification with large $\mu$ and a heavier
 gluino \cite{Joshipura:mu:2012}, viability of GUT-scale tribimaximal mixing 
 \cite{Joshipura:tb:2011}, and unified description of fermion
masses with quasi-degenerate (QD) neutrinos \cite{Joshipura:QD:2011} have been explored.
 A comparison of quality of different models has been
 also discussed \cite{Altarelli-Blankenburg:2011}. Recently
 existence of flavour symmetries \cite{Grimus:2016} and emergence of
 ordered anarchy from $5.{\rm dim.}$ theory \cite{Five dim:2015}, and Sparticle spectroscopy \cite{Fukuyama:2016} have
 been also investigated with numerical analyses on fermion masses. 
However, there exists a large class of SUSY SO(10) models where a
qualitative or at most a semi-quantitative representation of fermion masses
have been considered adequate without $\chi^2$ estimation. Examples from a very small part of a huge list are
\cite{Fukuyama:2004ps,mkp:2008,nofit0,nofit1,nofit2,nofit3,nofit4,Malinsky-Romao-Valle:2005,PSB-rnm:2010,Babu-Pati:2010,Fukuyama:Axion:2005,Romao:2015,Hisano:Zp:2016,Bobby:2016,Fukuyama:Infl:2005,Ellis:2016,Shafi:2016}. Even while
confronting other challenging problems through SUSY SO(10),  explanation of
 neutrino data only has been considered adequate; some examples out of many such
 works in this direction include derivation of new seesaw mechanism with TeV scale $Z'$
 \cite{Malinsky-Romao-Valle:2005}, prediction of Axions \cite{Fukuyama:Axion:2005}, low-mass $Z'$ induced by flavor
 symmetry \cite{Hisano:Zp:2016}, realization of SUSY SO(10) from
 ${\bf{\rm M\, -theory}}$ \cite{Romao:2015,Bobby:2016}, predictions
 of inflaton mass \cite{Fukuyama:Infl:2005}, and 
 Starobinsky type inflation \cite{Ellis:2016}, or quartic inflation
 \cite{Shafi:2016} from SUSY SO(10). Generalised hidden flavour
 symmetries have been explored without confining to any particular
 type of fermion mass fits \cite{Bajc-Smirnov:2016}.  

Despite many attractive qualities of SUSY GUTs including the
resolution of the gauge hierarchy problem, no experimental evidence of
supersymmetry has been found so far. This has led to search for gauge coupling unification of the standard gauge theory 
in non-supersymmetric (non-SUSY) GUTs  while sacrificing the elegant solution to
the gauge hierarchy problem in favour of fine-tuning to every loop
order \cite{Weinberg:2007,Barr:2010}. As stated above, single step
breakings of all popular non-SUSY GUTs including SU(5) \cite{su5} and
SO(10) \cite{so10} under the constraint of
the minimal fine-tuning hypothesis
\cite{del Aguilla:1981,rnm-gs:1983} fail to unify gauge
couplings.  

Introducing  gravity induced corrections through higher
dimensional operators \cite{ppm:1989} or additional fine-tuning of parameters with lighter
 scalars or fermions, gauge coupling unification in  non-SUSY SU(5) GUT has been
 implemented 
 \cite{MINGUTrefs,Bajc-gs:2006} including RH neutrino as DM \cite{Ma-Suematsu:2008}. Such unification has been also achieved  including triplet fermionic
 DM  \cite{Aizawa:2014}. A color octet fermion with mass $ > 10^8$
 ~GeV which is also needed for unification has
 been suggested as a source of non-thermal DM via non-renormalizable
 interactions \cite{Aizawa:2014}. As the model does not use matter parity \cite{Kadastic:dm1,Kadastic:dm2,Hambye:Rev,Frig-Ham:2010,mkp:2011,Hagedorn:2016}, the stabilising discrete
 symmetry for DM has to be imposed externally and appended to the GUT
 framework. Further, issues like neutrino masses and mixings and
 the baryon asymmetry of the universe have not been addressed in
 this model. Naturally the  non-SUSY SU(5) models\cite{ppm:1989,MINGUTrefs,Ma-Suematsu:2008,Aizawa:2014} have no
 explanation for the monopoly of parity violation in weak interaction
 alone \cite{Pati-Salam:1974,rnmjcp:1975}.

However, with or without broken D-Parity at the GUT scale
\cite{cmp:1984,cmgmp:1985}, non-SUSY SO(10) has been shown to unify
gauge couplings having one or more intermediate symmetries
\cite{cmp:1984,cmgmp:1985,rnm-mkp:1993,lmpr:1995,Deshpande:1993}. Extensive
investigations in such models have been reported with high
intermediate scales
\cite{Joshipura:2011,cmp:1984,cmgmp:1985,rnm-mkp:1993,lmpr:1995,Deshpande:1993,Berto:2009,Babu-Khan:2015,Babu-Bajc-Saad:2016,Meloni:2016,Buccella:2017}
and also with TeV scale $W_R,Z_R$ bosons and verifiable seesaw
mechanisms
\cite{ap:2012,pp:2013,app:2013,bpn-mkp:2013,pas:2014,PSB-rnm:2015,mkp-sahoo:NP:2016,mkp-bpn:2016,Matti:2014}. Out
of a large number of possible models that are predicted from non-SUSY 
SO(10) \cite{cmgmp:1985} fermion mass fit 
has been investigated only in one class of models with Pati-Salam intermediate symmetry
\cite{Joshipura:2011,Babu-Khan:2015,Meloni:2016} and also including additional
vector-like fermions \cite{Babu-Bajc-Saad:2016}.  The issue of DM
has been also
 addressed with different types of high scale intermediate symmetries
 and by introducing additional fermions or scalars beyond those needed
 by extended survival hypothesis  \cite{Mambrini:2015} but without addressing fermion mass fits. The problem of
 TeV scale $W_R$ boson prediction along with DM have been also addressed in non-SUSY SO(10) by invoking
 external $Z_2$ symmetry \cite{PSB-rnm:2015} without fitting
 charged fermion masses as also in a number of other  models
 \cite{cmgmp:1985,rnm-mkp:1993,Deshpande:1993,Berto:2009,Buccella:2017,ap:2012,pp:2013,app:2013,pas:2014,bpn-mkp:2013,Lindner:1995,Pilaftsis:1997,Arbelaez-Malinsky:2015}.
 As there has
been no experimental evidence of supersymmetry so far, likewise there
has been also no definite evidence of any new gauge boson beyond those of
the SM.
 This in turn has 
prompted authors to implement
 gauge coupling unification with the SM gauge symmetry below
the GUT scale 
\cite{MINGUTrefs,Bajc-gs:2006,Ma-Suematsu:2008,Kadastic:dm1,Kadastic:dm2,Frig-Ham:2010,mkp:2011,Aizawa:2014,Hagedorn:2016} by the introduction of additional
particle degrees of freedom with lighter masses. A natural question in this context is
how much of the advantages of the SUSY GUT paradigm is maintained in
the case of non-SUSY gauge coupling unification models. 
While SUSY SO(10) is
well known for its intrinsic R-Parity
\cite{Mohapatra:1986,SPMartin:1992} as gauged discrete symmetry 
\cite{Krauss:1989}  for
the stability of dark matter, as an encouraging
factor in favour of the non-SUSY GUT it has been shown
recently \cite{Kadastic:dm1,Kadastic:dm2,Hambye:Rev,mkp:2011,Hagedorn:2016} that matter parity defined as $P_M=(-1)^{3(B-L)}$ could be the corresponding discrete
symmetry intrinsic to non-SUSY SO(10) where $B (L)$ stands for
baryon (lepton) number. Whereas neutralino or wino
are predicted as dark matter candidates in SUSY GUTs, in non-SUSY
SO(10) the DM candidates could be non-standard fermions (scalars)
carrying even (odd) matter parity. In fact all SO(10) representations
have been identified to carry definite values of matter parity which
makes the identification of a dark matter candidate transparent from among the
non-standard scalar(fermion) representations. Thus there is enough  scope within non-SUSY SO(10) to implement the DM paradigm along with
an intrinsic stabilising symmetry. 

Compared to SUSY GUTs, the non-SUSY
GUTs do not have the problems associated with  the Higgsino mediated proton
decay \cite{Babu-Barr:1996,Babu-Pati:2010} while the canonical proton decay
mode $p\to e^+\pi^0$  has been accepted as the hall mark of predictions of
non-SUSY GUTs since more than four decades.
Further, the non-SUSY GUT also does not suffer from the well known gravitino
problem.\cite{Khlopov:1982,Majee:2008}.\\

Coupling unification in the single step breaking of
 non-SUSY SO(10) has been  addressed in an interesting paper by
 Frigerio and Hambye (FH) \cite{Frig-Ham:2010} by exploiting the intrinsic
 matter parity of SO(10) leading to 
 triplet fermion in ${45}_F$ as  dark matter candidate. The presence of a color octet
 fermion of mass $\ge 10^{10}$ GeV has been also noted for
 unification. The proton lifetime has
 been predicted in this model at two-loop level of gauge coupling
 unification. However details of fitting the neutrino oscillation data
including derivation of Dirac neutrino mass matrix and the
RH$\nu$ mass spectrum have not been addressed. Likewise related
details of derivation of
the baryon asymmetry of the 
universe via leptogenesis has been left out from the purview of
discussion. An added attractive aspect of the model is the discussion of
various methods, both renormalizable and non-renormalizable, by which
the triplet fermionic DM can have TeV scale mass.
Although proton lifetime has been predicted from the two-loop
determination of the GUT scale, important modification due to
threshold effects that could arise from the superheavy components of
various representations \cite{Weinberg:1980,Hall:1981,Ovrut:1981,mkp-pkp:1991,Langacker:1994} need further investigation. 

The contents of the present paper
 are substantially different from earlier works in many respects.
 We have discussed the matching with the neutrino oscillation data in
 detail where, instead of type-I seesaw, we have used hybrid seesaw  which is a combination of both type-I and type-II
  \cite{Akhmedov:2003dg}. Both of the seesaw mechanisms are
 naturally predicted in matter parity based SO(10) model having their
 origins rooted in the Higgs representation ${126}_H$  and the
 latter's coupling to the fermions in the spinorial representation
 $16$ through $f16.16.{126}_H^{\dagger}$.  Unlike a number of
neutrino mass models adopted earlier, in this work we have not assumed dominance of any one of the two seesaw mechanisms over the other.
  For the purpose of the present work we have determined the Dirac
  neutrino mass matrix at the GUT scale from the extrapolated
  values of charged fermion masses \cite{dp:2001} and exploiting the
  exact quark lepton symmetry \cite{Pati-Salam:1974} at that scale. With a view to
  investigating basis dependence of leptogenesis, the Dirac neutrino mass
  estimation  has been carried out in two ways: by using the $u$-quark
  diagonal basis as well as the $d$-quark diagonal basis.
Using these in the hybrid seesaw formula which is quadratic in the
Majorana coupling $f$ gives two distinct patterns of mass eigen values
for the heavy RH$\nu$ masses: (i)Compact scenario  where all masses are
heavier than the Davidson-Ibarra (DI) bound, and (ii) The hierarchical
scenario where only the lightest $N_1$ mass is below the DI bound.
  Thus each of these sets of RH neutrino masses
corresponds to two types of Dirac neutrino mass matrices or Yukawa
couplings which play crucial roles in the determination of
CP-asymmetry resulting from RH$\nu$ decays. We have carried out a
complete flavour analysis in determining the CP asymmetries. We have
also exploited solutions of Boltzmann equations in every case to
arrive at the predicted results on baryon asymmetry. Successful ansatz for
baryogenesis via leptogenesis is shown to emerge for each pattern of
RH$\nu$ masses. With the compact pattern of RH$\nu$ mass spectrum, this
 occurs when the
Dirac neutrino masses are determined in the $u$-quark or the $d$-quark
diagonal basis. However, in the hierarchical scenario of RH$\nu$
masses, the dominant CP asymmetry that survives the washout due to
$N_1$-decay and  contributes to the desired
baryon asymmetry is generated by the decay of the second generation
RH$\nu$ where the Dirac neutrino mass corresponds to the $u$-quark
diagonal basis. Because of the heavier mass
of the LH triplet scalar, although its direct decay to two leptons
\cite{Hambye-gs:2003} gives negligible contribution to the generated
CP-asymmetry, the additional vertex correction generated by its
mediation to the RH$\nu$ decay
 is found to lead to a
CP-asymmetry component comparable to other dominant contributions.      
Thus the same heavy 
triplet scalar $\Delta_L$ and the RH$\nu$s which drive the hybrid
seesaw formula for neutrino masses  and mixings are shown to generate
the leptonic $CP$ asymmetry leading to the experimentally observed
value for the baryon asymmetry of the universe over a wide range of
the parameter space in the model.\\ 

For the embedding of the suggested triplet fermionic DM
\cite{cirelli:2007xd} in  SO(10)
\cite{Frig-Ham:2010}, we assume it to originate from the non-standard
fermionic representation ${45}_F\subset SO(10)$ carrying even matter parity.
Having exploited the triplet fermionic DM $\Sigma_F(1,3,0)$ and the LH triplet Higgs scalar
$\Delta_L(1,3,-1)$ mediating the hybrid seesaw for neutrino masses and leptogenesis, we justify the presence
of these light degrees of freedom as ingredients for coupling
unification through their non-trivial contribution to the
$SU(2)_L\times U(1)_Y$ gauge coupling evolutions. In addition, we need
lighter scalar or fermionic octets with mass $\sim 5\times 10^7$ GeV
under $SU(3)_C$ to complete the precision gauge coupling unification. 

The degrees of freedom used in this model having their origins from
SO(10) representations ${126}_H, {10}_H, {45}_H$, and ${45}_F$ are
expected to contribute substantially to GUT threshold effects on the
unification scale through their superheavy components even without
resorting to make the superheavy gauge boson masses non-degenerate as
has been adopted in a number of earlier works for proton stability. 
It is important to note
that if we accept the stabilising symmetry for DM to be matter parity,
then the participation of ${126}_H\subset$ SO(10) in its spontaneous
symmetry breaking is inevitable. This in turn
dictates a dominant contribution to threshold effects 
on proton lifetime which has been ignored earlier but estimated in this direct breaking chain for the first
time. In addition the superheavy fermions in ${45}_F$ have been noted 
to contribute substantially. A possibility of partial cancellation 
of scalar and fermionic threshold effects 
is also pointed out. Although it is challenging to rule out the
present model by proton decay experiments, the predicted proton lifetime in this model 
for the $p\to e^+\pi^0$ is found to be within the accessible range of
the ongoing search limits \cite{Super K.,Hyper K.} for a wider range
of the parameter space.  

Unlike the case of direct breaking of SUSY SO(10) to MSSM
\cite{Joshipura:2011} or non-SUSY SO(10) through  Pati-Salam
intermediate symmetry \cite{Joshipura:2011}
, but like very large number of cases of model building in  non-SUSY GUTs, it is not our present goal to
  address charged fermion mass fit. But we discuss
  in Appendix C how all fermion masses  may be fitted at least
  approximately in future without substantially affecting this model predictions.

This paper is planned in the following manner.  In Sec.\,\ref{sec:mnu}
we discuss successful fit to the neutrino oscillation data where we
estimate the LH Higgs triplet and the
RH$\nu$ masses.
In Sec.\,\ref{sec:cp} we present the estimations of CP-asymmetry for different 
flavor states. In Sec.\,\ref{sec:be} we discuss Boltzmann equations
for flavour based analysis. In Sec.\,\ref{sub:n1} and Sec.\,\ref{sub:n2}
we present the results of final baryon asymmetry. In Sec.\,\ref{sec:dm} 
we discuss why the neutral component of fermionic triplet is a suitable 
dark matter candidate. In Sec.\,\ref{sec:uni} we discuss unification
of gauge couplings and determine the unification scale. In Sec.\,\ref{sec:th} we discuss 
 proton lifetime prediction including
GUT-threshold uncertainties. In Sec.\,\ref{sec:conc} we summarize and state conclusions. In 
Appendix A and Appendix B we provide renormalization group coefficients for gauge
coupling evolution and estimation of threshold effects. In Appendix C
we discuss the possibility of  parameterization of fermion masses.\\
 
\section{Hybrid Seesaw Fit to Neutrino Oscillation Data}\label{sec:mnu}
 In this section we address the issue of fitting the neutrino masses and mixings
 as determined from the  neutrino oscillation data by the hybrid
 seesaw formula. We then infer on the masses of heavy left-handed triplet
 and RH neutrinos necessary for leptogenesis.\\ 
 
After SO(10) breaking, the  relevant part of the  Lagrangian under SM
symmetry  is 
\begin{eqnarray}
{\cal - L}_{\rm Yuk}&\ni&Y^{ij}_\nu \bar{N}_{R_i}L_jh^\dagger
+\frac{1}{2}f^{ij}v_R N^T_{R_i}CN_{R_j} + \frac{1}{2}f_{ij} L^T_i
Ci\tau_2 \Delta_L L_j \nonumber \\
&& - \mu H^T i\tau_2 \Delta_L H +  M^2_\Delta Tr(\Delta_L^\dagger \Delta_L) 
+h.c.\label{Yukhiggs}
\end{eqnarray}
 The first term
  on the right-hand side (RHS) of eq.(\ref{Yukhiggs}) is from the SO(10)
  symmetric Yukawa term $Y^{(10)}.16.16.{10}_H$ whereas the second and
    the third terms are from $f.16.16.{126}^{\dagger}$ \cite{baburnm:1993}. 
    Also we have defined
 $v_R\equiv\left<\Delta_R \right> \sim M_{R}$ and $\mu=\lambda
v_R$. Although the associated RH scalar field $\Delta_R(1, 3, -2, 1) \subset
{126}_H$ has the respective quantum number under the LR gauge group
$SU(2)_L\times SU(2)_R\times U(1)_{B-L} \times SU(3)_C (\equiv
G_{2213})$, it is the singlet component $\Delta_R(1,1,0)$ under the SM
  that acquires the vacuum expectation value (VEV) $=v_R$. Similarly
  the LH triplet scalar field $\subset {126}_H$  has the transformation
  property $\Delta_L(3, 1, -2, 1)$ under $G_{2213}$ but the quantum
  numbers under the SM ($=G_{321}$)  are
  $\Delta_L(1, 3, -1)$.  Here $\lambda$ is the quartic
coupling of the SO(10) invariant Lagrangian resulting from the
combination of
 $10_H$ and $126_H$: $\lambda {10}_H^2.{126}^{\dagger}_H.{126}_H\supset
  \mu H^T i\tau_2 \Delta_L H $. The Higgs triplet mass-squared term has its
  origin from $  M^2_\Delta {126}^{\dagger}_H {126}_H$. 

Other notations are self explanatory.  
The hybrid formula  for the light neutrino mass matrix is the sum of
 type-I and type-II seesaw contributions \cite{rnm-gs:1983}
\begin{equation}
m_\nu=fv_L - M_D\frac{1}{fv_R}M^T_D, \label{eq:nu_mass}
\end{equation}
where $v_L={\lambda v_R v^2_{\rm ew}}/{M^2_\Delta}$ is the 
induced VEV of triplet scalar $\Delta_L$, and 
$M_D\equiv Y_\nu v_{\rm ew}$. 

There is the well known standard ansatz to fit fermion masses in
SO(10) along the line of \cite{baburnm:1993}.
To estimate the Dirac mass matrix in this work 
we have carried out one-loop renormalization group evolution of  
Yukawa couplings in the bottom-up approach using PDG values of all
charged fermion masses. At the electroweak scale $\mu=M_Z$ using experimental 
data on charged fermion masses we choose up-quark  or down-quark mass diagonal bases 
in two different  scenarios. We then evolve them upto the GUT scale $\mu=M_{U}$
using bottom-up approach \cite{dp:2001}. At this scale we assume equality of 
the up-quark and the Dirac
neutrino mass matrices, $M_D \simeq M_u$, which holds upto a very good
approximation in SO(10) due to its underlying quark-lepton symmetry
\cite{Pati-Salam:1974}.\\

As pointed out in Sec.1, $\chi^2$ fit to all
fermion masses and mixings in SUSY SO(10) or in non-SUSY SO(10) with $G_{224}$
intermediate symmetry requires a small departure from this assumption
\cite{Joshipura:2011,Babu-Khan:2015,Meloni:2016}.
On the other hand a very recent derivation of neutrino mass and mixing sum-rules has
been found to require  $M_D$ close to $ M_u$
\cite{Buccella:2017} as in our case. Although in the present case of
non-SUSY SO(10) breaking directly to the SM gauge theory, fermion mass
fit is not our goal in this paper, we have discussed the issue
in Appendix C.\\

 We further assumed that $M_D(M_{M_{GUT}}) \sim M_D(\mu)$ for 
all lower mass scales $\mu <M_{GUT}$. We could have done better to estimate 
the Dirac mass matrix at the electroweak scale by following the top-
down approach  but since it does not 
get appreciable correction 
due to the absence of the strong gauge coupling $\alpha_{3C}$
 \cite{dp:2001} contribution, this approximation does not influence our final result 
 substantially. Another reason is that for leptogenesis we need 
 Dirac neutrino Yukawa couplings at intermediate scales, $\mu \sim
 (10^6-10^{12})$ GeV where the renormalisation group (RG) running effects are
 expected to be smaller in the top-down approach. 

 Thus in the down quark diagonal basis under the assumption of negligible RG effects we have  at $\mu=M_Z$ 
 
\begin{equation} 
M^{(d)}_D({\rm GeV})=\begin{pmatrix}
0.01832+0.00441i & 0.08458+0.01114i & 0.65882+0.27319i\\
0.08458+0.01114i & 0.38538+1.56\times10^{-5}i& 3.32785+0.00019i\\
0.65882+0.27319i & 3.32785+0.00019i & 81.8543-1.64\times10^{-5}
\end{pmatrix}\label{eq:MD_ddia}
\end{equation}
 We repeat the above procedure in the 
 up-quark diagonal basis at $\mu = M_Z$ instead of the down quark diagonal
 basis leading to    
\begin{equation} 
M^{(u)}_D({\rm GeV}) =\begin{pmatrix}
0.00054 & (1.5027+0.0038i)10^{-9} &  (7.51+3.19i)10^{-6} \\
(1.5027+0.0038i)10^{-9}& 0.26302 &  9.63\times 10^{-5} \\
(7.51+3.19i)10^{-6} & 9.63\times 10^{-5} & 81.9963 \\
\end{pmatrix}.\label{eq:MD_udia}
\end{equation}      

For the sake of clarity it might be necessary to explain how the mass matrix
structure given in eq.(\ref{eq:MD_udia}) emerges with very small non-diagonal elements. 
 In the bottom-up approach for the RG evolution of Yukawa matrices, we have assumed the up-quark mass matrix $M_u(M_Z)$ to be diagonal in one case at the electroweak scale which we designate as up-quark diagonal basis. In this case
 naturally
all elements of the down quark mass matrix $M_d(M_Z)$ are non-vanishing. In the alternative case,
called the d-quak diagonal basis, we have chosen $M_d(M_Z)$ diagonal
for which all nine elements of $M_u(M_Z)$ are non-vanishing. 
In the case of up-quark diagonal basis, however, the non-diagonal elements of 
$M_u(M_{M_{GUT}})$ acquire non-vanishingly small corrections due to RG effects in the bottom-up approach and this is approximated as the Dirac-neutino mass matrix
$M_D^{(u)}(M_{GUT})$. This explains the appearance of non-diagonal elements appearing in eq.(\ref{eq:MD_udia}). It may  be noted further that the RG-corrections in the Dirac neutrino mass matrix $M_D^{(u)}$ for evolutions from $\mu=M_{GUT}$ down to relevant lower scales have been ignored as they are expected to be much smaller.\\ 

The Dirac neutrino mass matrices given in eq. eq.(\ref{eq:MD_ddia})
and eq.(\ref{eq:MD_udia}) 
 are used in the second term of the right-hand side (RHS) of eq.(\ref{eq:nu_mass})
where in the left-hand side (LHS) we use the value of light neutrino
mass matrix for the normally ordered case  with $m_{\nu_1}=0.00127$\,eV and the best fit values for other parameters
\cite{Capozzi:2013csa}. We have also assumed that Majorana phases are zero at all mass scales. 
      
\begin{figure}[h!]
\centering 
\includegraphics[scale=0.78]{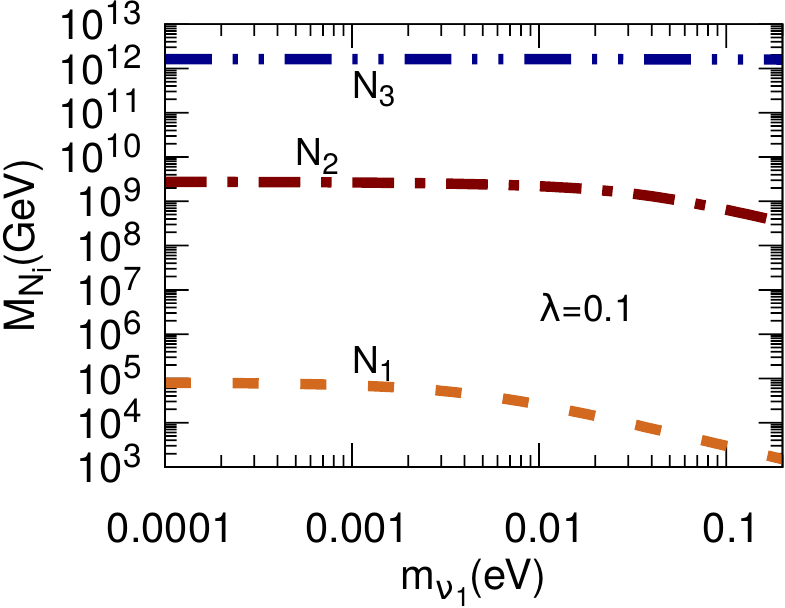}
\includegraphics[scale=0.78]{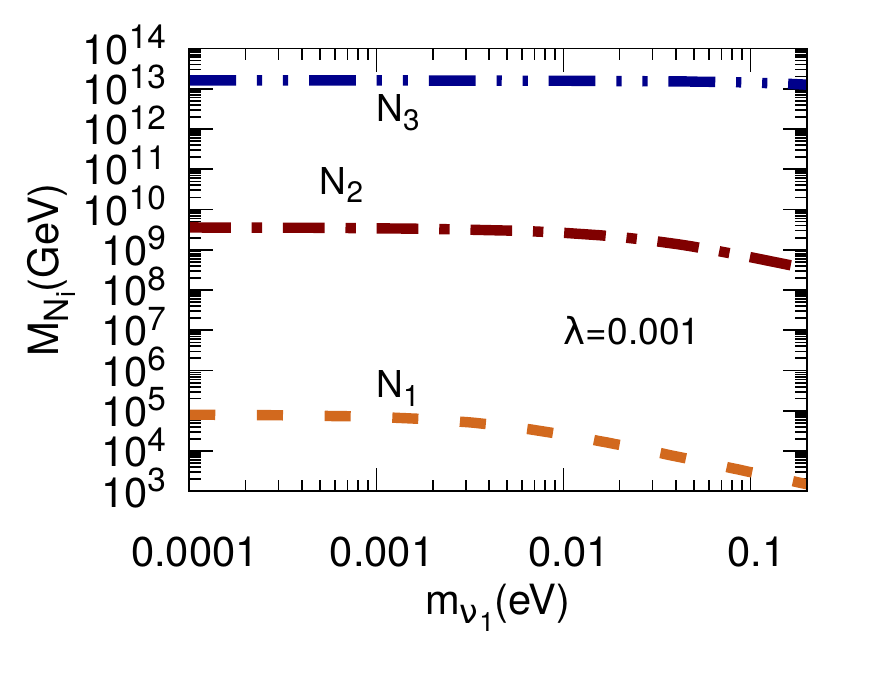}\\
\includegraphics[scale=0.79]{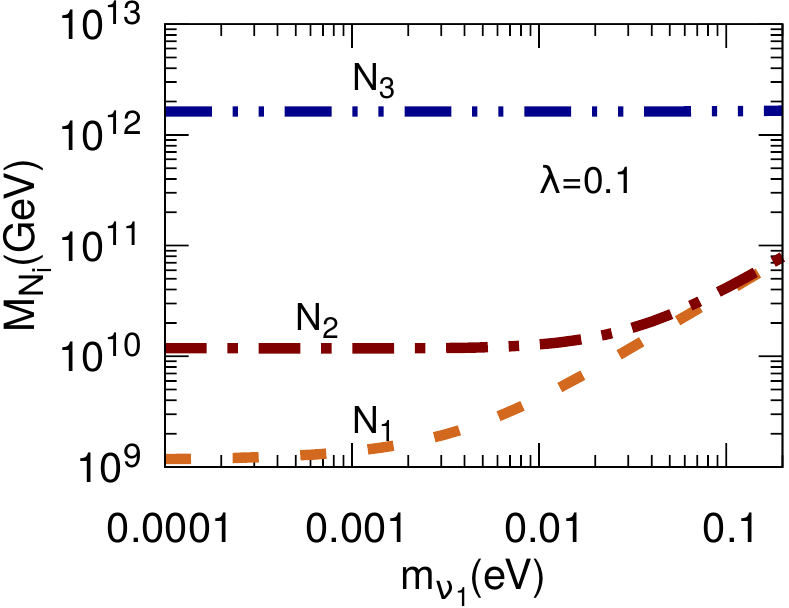}
\includegraphics[scale=0.79]{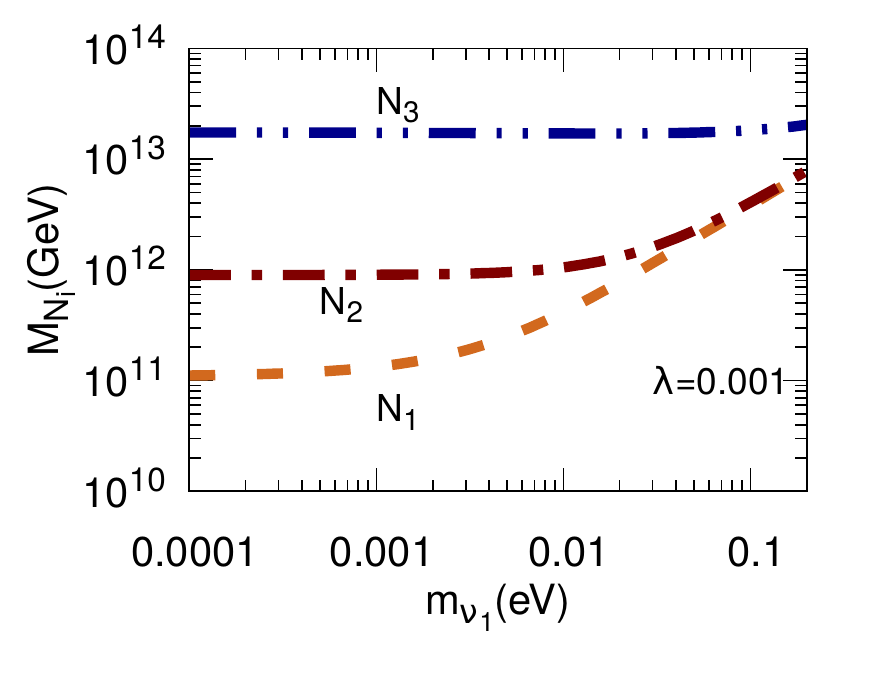}
\caption{Prediction of heavy RH neutrino masses as a function of the
  lightest neutrino mass and the quartic coupling $\lambda$
in the case when the three neutrino masses are normally ordered. The top row 
represents a hierarchical spectrum solution of RH neutrinos and the bottom row represents a 
 not so hierarchical scenario which we call as compact spectrum solution . The values of $M_{\Delta_L}=10^{12}$\,GeV
and $v_R=10^{15.5}$\,GeV have been kept fixed. The value of the quartic coupling 
used here has been taken to be $\lambda=0.1(0.001)$ for the left panel (right panel).}
\label{fig:hcmass}
\end{figure}

We then search for solutions for the Majorana coupling $f$ or,
equivalently, the values of RH neutrino masses.       
Due to strongly  hierarchical structure of $M_D$ matrix, it is impractical to 
assume the dominance of the type-I or the type-II term in the hybrid
seesaw formula of eq.\,(\ref{eq:nu_mass}). Since  eq.\,(\ref{eq:nu_mass}) 
is quadratic in $f$,  
it has two solutions for every eigenvalue and thus giving a total of $2^3=8$
plausible solutions \cite{Abada:2008gs}. But for a given $M_D$ and $m_\nu$ there should
be only two distinct positive definite solutions. We estimated these solutions 
for $f$  using
the neutrino oscillation  data of ref.\cite{Capozzi:2013csa} as input and numerical iteration.  
A robust iterative numerical estimation of $f$ matrix is performed to match 
the  oscillation data. 
Thus by fixing the lightest neutrino mass and the  VEV $v_L$ in a chosen hierarchy of light neutrino 
masses, the precise forms 
of the two solutions with positive definite $f$ are evaluated upto the desired precision. These solutions are presented in Fig.\,\ref{fig:hcmass} for two sets of values of quartic coupling, $\lambda=0.1$ and $\lambda=0.001$.
%%%%%%%%%%%%%%%%%%%%%%%%%%%%%%%%%%%%%%%%%%%%%%%%%%%%%%%%%%%%%%%%%%%
%%%%%%%%%%%%%%%%%%%%%%%%%%%%%%%%%%%%%%%%%%%%%%%%%%%%%%%%%%%%%%%%%

%\begin{figure}
%\centering 
%\includegraphics[scale=0.8]{mfile1.pdf}
%\includegraphics[scale=0.8]{mfile2.pdf}\\
%\includegraphics[scale=0.8]{mfile3.pdf}
%\includegraphics[scale=0.8]{mfile4.pdf}
%\caption{Right handed neutrino masses vs lightest neutrino 
%in the normal hierarchy of active neutrinos. The left panel
%corresponds to d-quark diagonal basis and the right panel to 
%u-quark diagonal basis at $M_Z$. We call it compact 
%scenario.}\label{fig:cmass}
%\end{figure}
\newpage
In Fig.\,\ref{fig:hcmass} we have presented these solutions for 
the normally ordered
values of active light neutrino masses. Solutions in the top row of the 
figure have strongly hierarchical heavy RH neutrino masses, lightest of them being 
$M_{N_1}\sim {\cal O}(10^{3-5})$\,GeV, testable in future collider experiments,
and the heaviest $M_{N_3}\sim {\cal O}(10^{12})$\,GeV. We call such solutions
of RH neutrino masses to represent a hierarchical spectrum scenario. 
Solutions in the bottom row of the figure are not so hierarchical and the RH 
neutrinos only span three orders of magnitude of mass range. We call the solutions of this type given in the bottom row to represent a compact spectrum scenario.  
Lightest of RH neutrino in this scenario is 
$\sim {\cal O}(10^{9-11})$\,GeV which is  far away from direct detection limit of any collider experiment.
In arriving at these solutions we assumed the LH triplet scalar mass $M_{\Delta_L}=10^{12}$\,GeV, GUT symmetry breaking 
VEV $v_R=10^{15.5}$\,GeV, and the value of the quartic coupling $\lambda=0.1\,(\rm left\,panel)$ and $0.001\,(\rm right\, panel)$. 
We note that the RH$\nu$ masses increase with decrease in $\lambda$ for the compact spectrum scenario while
it almost stays unaffected in the hierarchical spectrum  scenario. Also the theory should 
continue to remain perturbative on acquiring $N_1$-dominated
leptogenesis  because increasing
$\lambda (\sim 1)$ for 
the above value of $M_\Delta$ will make $M_{N_1}<10^9$\,GeV and  $N_1$- dominated 
leptogenesis will not be possible.

In the compact spectrum scenario we estimate  the $f$ matrix in the $d$-diagonal
basis using eq.(\ref{eq:MD_ddia}),
$m_{\nu_1}=0.00127$ eV, $M_{\Delta_L}=10^{12}$\,GeV,
and $v_R=10^{15.5}$\,GeV\\
\par\noindent{\large \bf{\underline {$M_D=M^{(d)}_D$}}}\\
\begin{equation}
f=\begin{pmatrix}
	0.385+0.1291i  &  0.4617-0.4922i  &  3.509+1.080i \\
	0.4617-0.4922i &  4.626+0.1567i &   22.80+0.3317i \\
    3.509+1.080i &   22.80+0.3317i  &  511.6+0.47i
\end{pmatrix}\times 10^{-6}.
\label{feqcd}
\end{equation}
For the same parameters in the compact spectrum scenario but with  $ M^{(u)}_D$ in
$u$-diagonal basis given in  eq.(\ref{eq:MD_udia}), we derive
\par\noindent{\large \bf{\underline {$M_D=M^{(u)}_D$}}}\\
\begin{equation}
f=\begin{pmatrix}
0.3175+0.0904i   & 0.1232-0.6089i  &  -0.4869-0.6918i \\
0.1232-0.6089i   & 3.610-0.0724i  &  1.587+0.2599i \\
-0.4869-0.6918i  &  1.587+0.2599i &  511.8+0.6524i
\end{pmatrix}\times 10^{-6}.
\label{feqcu}
\end{equation}
 In the hierarchical spectrum scenario, similarly, we have the two matrices for
 $f$ 
 \par\noindent{\large \bf{\underline {$M_D=M^{(d)}_D$}}}\\
\begin{equation}
f=\begin{pmatrix}
-0.0690+0.0147i & -0.341+0.0164i & -4.0194+1.5783i \\
-0.341+0.0164i & -1.5745-0.2133i & -20.2464-0.3306i \\
-4.0194+1.5783i & -20.2464-0.3306i & -507.895-0.4034i
\end{pmatrix}\times 10^{-6},
\label{feqhd} 
\end{equation}
\par\noindent{\large \bf{\underline {$M_D=M^{(u)}_D$}}}\\
\begin{equation}
f=\begin{pmatrix}
-0.000025+0.000008i & -0.00019-0.00215i & -0.00538-0.00177i\\
-0.00019-0.00215i & -0.56091+0.0092i & 0.95702-0.27084i \\
-0.00538-0.00177i & 0.95702-0.27084i & -508.16-0.60957i
\end{pmatrix}\times 10^{-6}.
\label{feqhu}
\end{equation}
Despite widely varying magnitudes of different elements in the matrix, 
the mass eigenvalues in the $u-$ quark and $d-$ quark diagonal bases are not very 
different in both the compact spectrum and the hierarchical spectrum scenarios. Therefore, 
we have presented only one set of solutions for the RH$\nu$ masses  in
Fig.\,\ref{fig:hcmass}. 
%Not only the eigenvalues but also the signs of elements of $f$ matrix
%guide us to call the compact scenario as type-II-like and 
%hierarchical scenario to be type-I-like. Another interesting
%observation is that a large VEV like $V_R\sim M_5$ and a value of LH
%Higgs triplet $\Delta_L$ mass nearly $3-4$ orders smaller are able to contribute
%reasonably to neutrino oscillation data via hybrid seesaw formula.
It is quite encouraging to note that despite the GUT scale value of
$v_R$, the type-II term does not upset the type-I seesaw term in the
hybrid formula, rather both of them contribute significantly to the
light neutrino mass matrix. We will explore the plausibility of
sufficient leptogenesis using the hybrid seesaw mechanism of this model to explain BAU.

\section{Baryon Asymmetry of the Universe}\label{sec:basy}
In this section at first we estimate the leptonic CP- asymmetry
generated in decays of both
RH$\nu$ and $\Delta_L$. The dynamically generated lepton asymmetry gets converted into
baryon asymmetry due to sphaleron interaction \cite{Sphaleron}. Leptogenesis is discussed in various papers \cite{leptogenesis}.
 The flavour independent 
calculation of asymmetry is applicable at high temperatures when all the charged lepton 
mediated interactions are out of equilibrium i.e. $T \gtrsim 10^{12}$\,GeV.
Flavour dependent analysis \cite{Barbieri:1999} becomes necessary for leptogenesis at
lower temperatures. In hierarchical spectrum scenarios we have
$M_{N_1}\sim 10^{3-5}$\,GeV which violates the Davidson-Ibarra bound \cite{DIbound} badly,
therefore it can not produce required amount of flavour independent lepton asymmetry. Instead it 
washes out the asymmetry produced at the early stage in $N_{2,3}$ decays. In the recent 
studies \cite{Barbieri:1999,Bari:2005,Vives,Bertuzzo:2010et,Antusch:2010ms} it has been shown that under such circumstances
the next heavy neutrino $N_2$ can produce the required asymmetry,  if 
$M_{N_2}\gtrsim 10^{10}$\,GeV and there exists a heavier $N_3$. If the asymmetry
produced by $N_2$ is not completely washed out by lightest neutrino $N_1$, it survives
and gets converted to baryon asymmetry. 
%We will calculate the surviving asymmetry
%for the GUT-scale constrained Dirac neutrino mass matrix in this scenario. 
On the other hand, in the compact spectrum scenario, the lightest RH neutrino is
well within the Davidson-Ibarra bound, therefore the asymmetry can be 
produced in the lightest RH$\nu$ decay. Since for a large region of
the parameter space
we have shown that $M_{N_1}<<10^{12}$\,GeV, the asymmetry will depend on 
flavour dynamics.

\subsection{CP- Asymmetry}\label{sec:cp}
\begin{figure}[h!]
\begin{center}
\includegraphics[scale=.4]{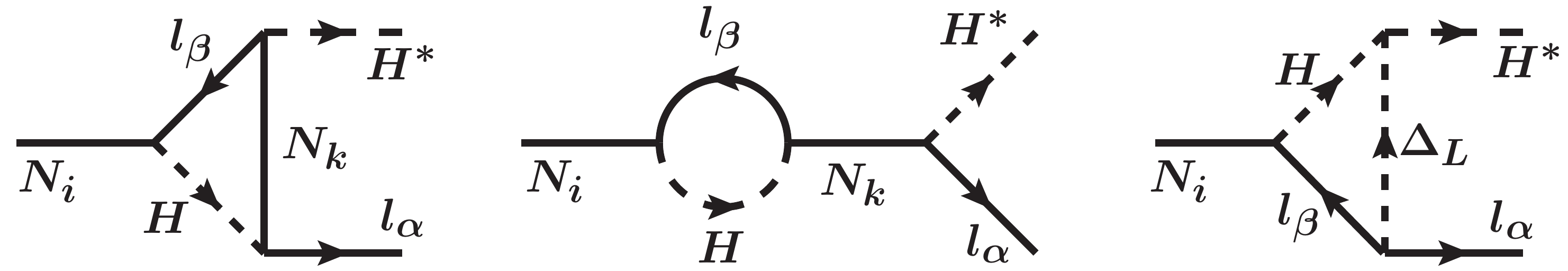}
\end{center}
\caption{ One-loop Feynman diagrams for the decay of  RH neutrino $N_i$.
The first and the third diagrams represent vertex corrections 
and the second diagram represents self-energy correction.}
\label{fig:feyn1}
\end{figure}

The flavoured CP-asymmetry in the decay of $N_i$ to a lepton $l_\alpha$ 
 is generated in the lepton flavor generation $\alpha$, 
 and is defined as \cite{Fong:2013wr,Covi:1996wh,Buccella:2012kc} 
\begin{equation}
\varepsilon_{i\alpha}=\frac{\Gamma (N_i \rightarrow l_\alpha + H^*) - 
\Gamma (N_i \rightarrow \bar{l}_\alpha + H)}{\sum_\beta
\left[ \Gamma (N_i \rightarrow l_\beta + H^*) +
\Gamma (N_i \rightarrow \bar{l}_\beta + H)\right]} \,.
\end{equation} 
One loop decay contributions of $N_i$ are mediated by either $N_{k\neq i}$ or $\Delta_L$
%The two distinct contributions to the asymmetry arise from $N_{k\neq i}$ mediation and 
%$\Delta_L$ mediation at one loop 
\cite{Hambye-gs:2003} as shown in Fig.\,\ref{fig:feyn1}. 
The total asymmetry is sum of the two contributions
\begin{equation}
\varepsilon_{i\alpha}=\varepsilon^N_{i\alpha}+\varepsilon^\Delta_{i\alpha}.
\end{equation}
%%%%%%%%%%%%%%%%%%%%%%%%%%%%%%%%%%%%%%%%%%%%%%%%%%%%%%%%%%%%%%%%%%%%
\begin{figure}[h!]
\centering 
\includegraphics[scale=0.4]{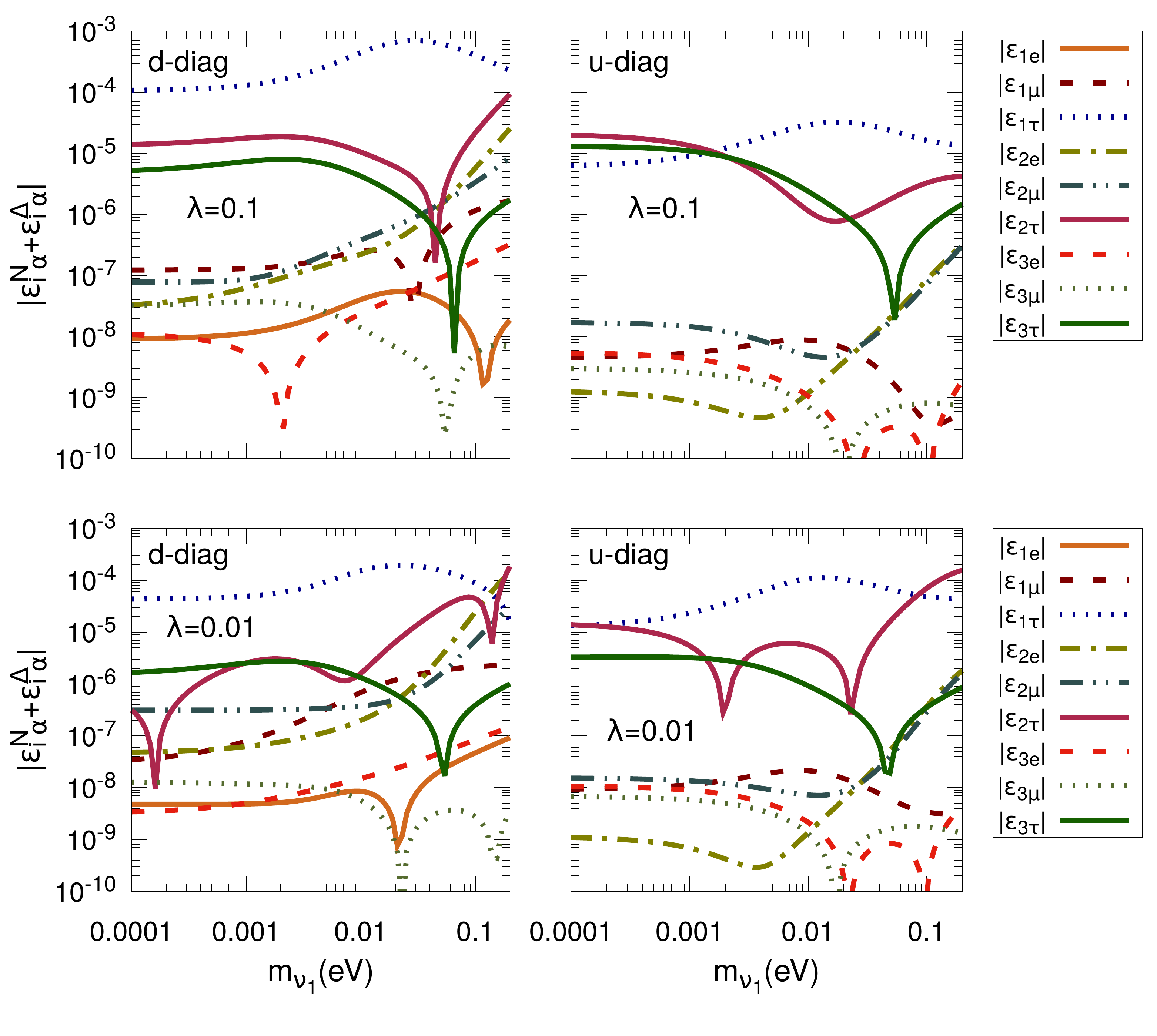}
\caption{The CP-asymmetry vs. lightest neutrino mass in the compact
  spectrum scenario.
The top left(right)-panel correspond to $d(u)$-quark diagonal basis
for $\lambda=0.1$. The bottom left(right) panel correspond to $d(u)$-quark
diagonal basis but for $\lambda=0.01$.}
\label{fig:cpac}
\end{figure}
The asymmetry produced in the $N_i$  decay due to $N_{k\neq i}$ appearing 
in the loop is \cite{Covi:1996wh, Buccella:2012kc}
\begin{eqnarray}
\varepsilon^N_{i\alpha}&=&\frac{1}{8\pi}\sum_{k\neq i}
\frac{{\rm Im} \left[\left(\hat{Y}_\nu^\dagger \right)_{i\alpha}
\left(\hat{Y}_\nu \right)_{\alpha k} \left(\hat{Y}_\nu^\dagger 
\hat{Y}_\nu \right)_{ik}\right]}{\left(\hat{Y}_\nu^\dagger 
\hat{Y}_\nu \right)_{ii}}~h\left(\frac{M_{N_k}^2}{M_{N_i}^2}\right) \nonumber \\
&+&\frac{1}{8\pi }\sum_{k\neq i}
\frac{{\rm Im} \left[\left(\hat{Y}_\nu^\dagger \right)_{i\alpha}
\left(\hat{Y}_\nu \right)_{\alpha k} \left(\hat{Y}_\nu^\dagger 
\hat{Y}_\nu \right)_{ki}\right]}{\left(\hat{Y}_\nu^\dagger 
\hat{Y}_\nu \right)_{ii}}~g\left(\frac{M_{N_k}^2}{M_{N_i}^2}\right)
\end{eqnarray}
The first line of this expression contains lepton number 
violating terms while the second line is the lepton number conserving but 
violates lepton flavour. Here, $\hat{Y}_\nu=Y_\nu U^*_f$ is the Dirac 
Yukawa coupling in the right-handed neutrino diagonal mass basis and 
$U_f$ is the unitary matrix diagonalizing $f$. The loop functions in the 
asymmetry expression are \cite{Buccella:2012kc}
\begin{eqnarray}
g(x)&=&\frac{1-x}{(1-x)^2+\left(\frac{\Gamma_i}{M_i}
-x\frac{\Gamma_k}{M_k}\right)^2} \nonumber \\
h(x)&=&\sqrt{x}\left[g(x)+1-(1+x){\rm log}\left(\frac{1+x}{x}\right)\right] .
\label{gheq}
\end{eqnarray}
Here by retaining the Wigner-Eckart term in the loop function we can handle
degenerate RH$\nu$ mass scenario without hitting singularity, 
%so that the function does not become singular in the degenerate heavy mass scenario, 
which is possible in compact spectrum scenario in our model (see Fig. \ref{fig:hcmass}).
Note that in the degenerate regime CP asymmetry gets largest contribution from
self-energy term and may reach to a value of ${\cal O}(1)$. 
%The first line of this expression is due to lepton number violation while the second
%line is due to lepton flavor violation but is lepton number conserving. 
%Here, $\hat{Y}_\nu=Y_\nu U^*_f$ is the Dirac Yukawa coupling 
%in the right-handed neutrino diagonal mass basis and $U_f$ is the unitary matrix
%diagonalizing $f$. Also the associated loop functions are
% , we have 
%kept the Wigner-Eckart like term in the self energy correction to one loop decay rate.
%This condition may enhance the $CP$-asymmetry. 
%In the region of resonance, the asymmetry can be of ${\cal O}(1)$.
%\beq
%1-x=\pm \left(\frac{\Gamma_i}{M_i}
%-x\frac{\Gamma_k}{M_k}\right)
%\eeq
%\begin{equation}
%g(x)=0.5\times {\left(\frac{\Gamma_i}{M_i}
%-x\frac{\Gamma_k}{M_k}\right)^{-1}}
%\end{equation}
\begin{figure}
\centering 
\includegraphics[scale=0.4]{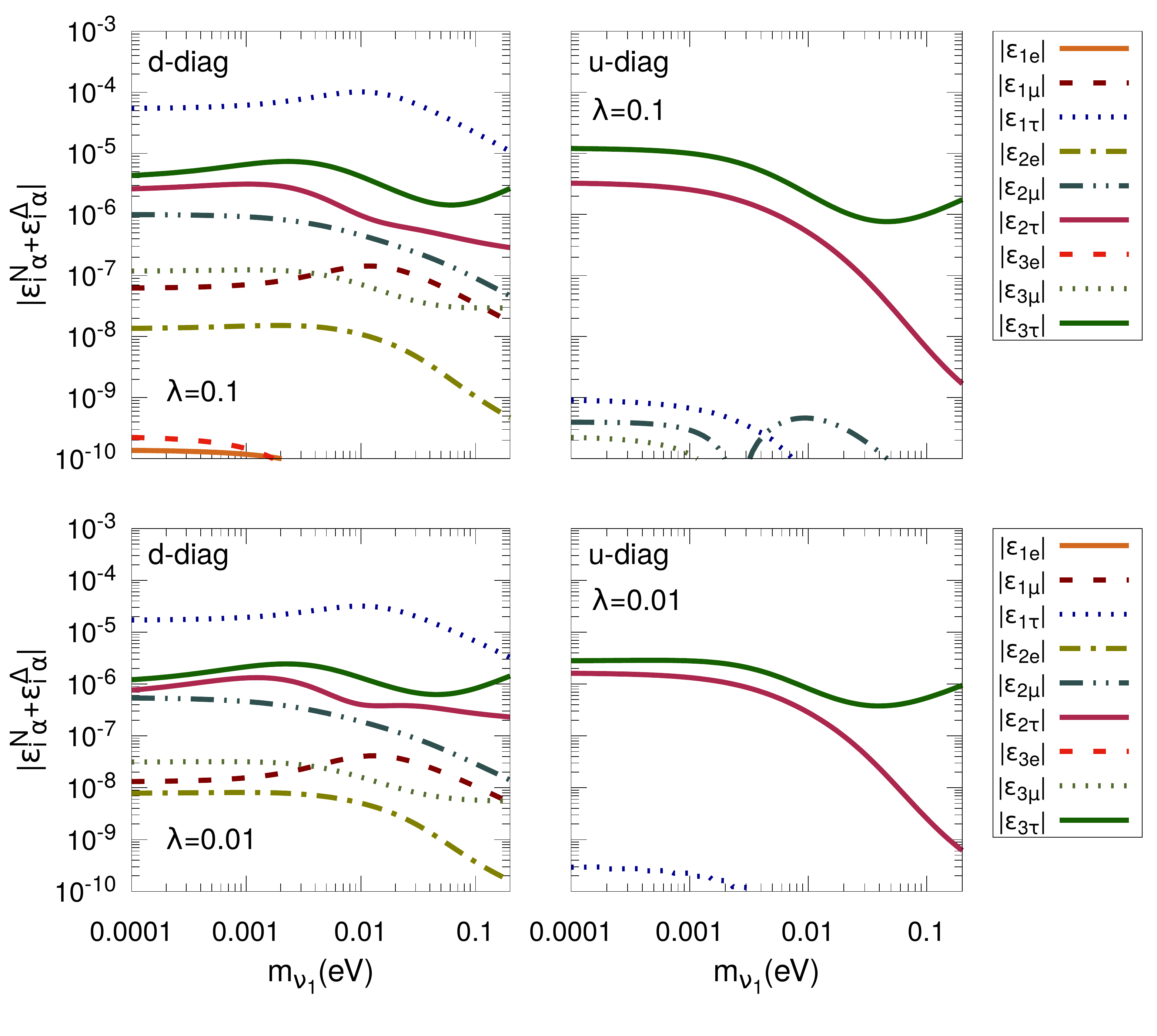}
\caption{The CP-asymmetry vs. the lightest neutrino mass for
  hierarchical spectrum scenario  of RH$\nu$ masses.
The top left (right)-panel correspond to d(u)-quark diagonal basis
for $\lambda=0.1$. The bottom left(right) panel correspond to d(u)-quark
diagonal basis but for $\lambda=0.01$.}
\label{fig:cpah}
\end{figure}
The $CP$-asymmetry produced in $N_i$ decay from the $\Delta_L$ mediated 
diagram is \cite{Hambye-gs:2003} 
\begin{equation}
\varepsilon^\Delta_{i\alpha}=-\frac{1}{4\pi}\sum_{\beta}
\frac{{\rm Im}\left[\left(\hat{Y}_\nu\right)_{i\beta} f^*_{\beta\alpha} 
\left(\hat{Y}_\nu\right)_{i\alpha}\mu \right] }
{\left(\hat{Y}^\dagger_\nu \hat{Y}_\nu\right)_{ii}M_{N_i}}
\left[1-\frac{M^2_\Delta}{M^2_{N_i}} 
\log\left(1+\frac{M^2_{N_i}}{M^2_\Delta}\right) \right],
\end{equation}
which gets contribution proportional to the trilinear coupling mass 
term $\mu$. Its loop function is larger for smaller $M_{\Delta_L}$.
But $M_{\Delta_L}$ can not be made arbitrarily small without decreasing 
$\mu$ or increasing $v_L$ 
%which will reduce CP-asymmetry linearly or without increasing $v_L$ 
which is constrained to be below GeV from electroweak (EW) 
precision constraints. Decreasing $\mu$ would decrease CP asymmetry linearly.

Keeping the GUT scale value of $v_R=10^{15.5}$\,GeV and $M_{\Delta_L}=10^{12}$\,GeV 
we have estimated the flavored CP-asymmetry for different values of the  lightest
neutrino mass in the normally ordered hierachical case of light
neutrino masses. Change in the mass of $m_{\nu_1}$
alters $f$ and thus changes the masses and mixings of RH$\nu$s.   
Flavour asymmetries for $N_i$ decay into $\alpha$ 
flavour are shown in Fig.\,\ref{fig:cpac} for compact spectrum case and in 
Fig.\,\ref{fig:cpah} for the hierarchical spectrum case of RH$\nu$s.
We note that variation in quartic coupling changes CP-asymmetry significantly, 
particularly in the hierarchical spectrum scenario. The tree level decay widths are
unaffected by the presence of the scalar triplet $\Delta_L$ in the scheme. 
%presence. and are 
%\begin{equation}
%\Gamma_{N_i}= \frac{1}{8 \pi} M_{N_i} 
%\left(\hat{Y}_\nu\hat{Y}^\dagger_\nu\right)_{ii} \,.
%\label{gammaN}
%\end{equation}
%Usually, in a large class of such theories type-I seesaw appears together
%with type-II seesaw. 

\begin{figure}[h!]
\begin{center}
\includegraphics[scale=.4]{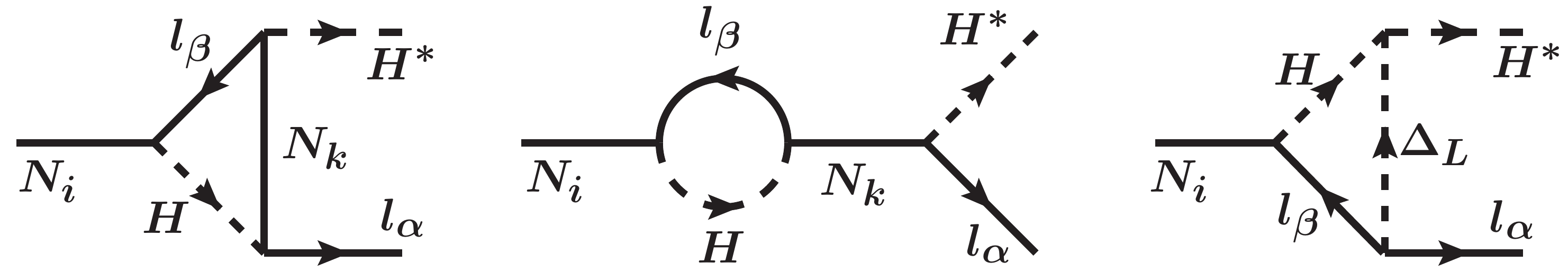}
\end{center}
\caption{Left handed triplet scalar one-loop decay.}
\label{fig:feyn2}
\end{figure}

The presence of the heavy scalar triplet $\Delta_L$ in our theory adds another
source of $CP$-asymmetry ($\epsilon_\Delta$) which is produced by the
decay of the 
triplet scalar itself into two like-sign or neutral leptons \cite{Hambye-gs:2003}.
Though one triplet scalar is enough to generate the active neutrino 
masses and mixings through type-II seesaw, the asymmetry production 
in $\Delta_L$ decay needs either more than one triplet scalars 
\cite{Hambye:2000ui,Ma:1998dx,Sierra:2014tqa,Donell} or combination 
of triplet scalar and right-handed neutrinos \cite{Hambye-gs:2003} as shown in
Fig.~\ref{fig:feyn2} for our model. The CP-asymmetry generated due to $\Delta_L$ 
decay and mediated by RH$\nu$ is written as \cite{Hambye-gs:2003}
\begin{eqnarray}
\varepsilon_\Delta &=& 2\cdot
{{\Gamma (\Delta_L^* \rightarrow l + l) - \Gamma (\Delta_L \rightarrow 
\bar{l} + \bar{l })}\over{\Gamma (\Delta_L^* \rightarrow l + l) 
    +\Gamma (\Delta_L \rightarrow \bar{l} + \bar{l}     )}}\nonumber \\
&=& \frac{1}{4 \pi} \sum_k M_{N_k} \frac{\sum_{il}{\cal I}m[(Y_\nu^\ast)_{ki} 
(Y_\nu^\ast)_{kl}
f_{il} \mu^\ast]}{\sum_{ij}|f_{ij}|^2
 M^2_\Delta + 4 |\mu|^2}
\log(1+M^2_\Delta/M^2_{N_k}) \,.
\label{eq:epsD}
\end{eqnarray}
We note that, since $v_R\simeq 10^{15.5}$\,GeV and $M_{\Delta_L}\simeq 10^{12}$\,GeV,
either of the two terms in the denominator of $\varepsilon_\Delta$ is large enough to
keep the $CP$-asymmetry fairly small for the parameters under consideration. 
For example, if three right-handed neutrino 
masses are $M_{N_k}=(6.6990, 13.869, 1431)\times 10^{9}$\,GeV, 
%their decay rates are estimated as 
% $\Gamma_{N_k}=(4.1437\times 10^3, 1.8661\times 10^5, 1.1509\times 10^{10})$\,GeV . 
%Correspondingly the three
% Hubble constants are  
% $H(T=M_{N_k})=(6.2820\times 10^1, 2.6927\times 10^2, 2.8665\times 10^6)$\, GeV.
% The decay rate for triplet scalar $\Gamma_\Delta=3.97887\times 10^{15}$ GeV with
%the Hubble constant
%$H(T=M_\Delta)=1.39985\times 10^6$ GeV.
the three CP-asymmetries due to $N_k$ decays from the first two diagrams of 
Fig. \ref{fig:feyn1} are
$|\epsilon_{N_k}|=  (4.7\times 10^{-5},5.1\times 10^{-8},
1.7\times 10^{-8})$. Likewise the CP-asymmetries from the third diagram are: 
$|\epsilon^\Delta_{N_k}|=(5.2\times 10^{-5}, 4.5\times 10^{-8}, 
2.4\times 10^{-6})$. Compared to these numbers, the CP-asymmetry due to $\Delta_L$ decay
of Fig. \ref{fig:feyn2} is $|\epsilon_\Delta| =2.1\times 10^{-12}$.
Also, since $M_{\Delta_L}>>M_{1,2}$, the asymmetry generated at the early 
stage will be washed out at the production phase of lighter RH$\nu$s.
Henceforth, we will ignore the $\Delta_L$ asymmetry in our numerical 
estimations \cite{Sierra:2014tqa}. In the next subsection we will estimate the 
lepton asymmetry using Boltzmann  equations for the system.

%In the scenario when there are more than one scalar triplets, 
%the self-energy diagrams for scalar triplet and vertex corrections mediated
%by triplet will contribute. The bounds on the lightest right-handed neutrino 
%masses in all such scenarios is of same order or larger as quoted earlier, and
%the left-handed triplet scalar is bounded to be $M_{\Delta_L}\gtrsim 10^{12}$\,GeV.

\subsection{Boltzmann Equations} \label{sec:be}
The evolution of number density is obtained by solving the set of 
Boltzmann equations. The co-moving number density is $Y_X\equiv n_X/s$.
The Boltzmann equations for heavy neutrinos number density are \cite{Abada:2008gs}
\begin{eqnarray}
\frac{dY_{N_i}(z)}{dz}&=&-K_i(D_i(z)+S_i(z))\left(Y_{N_i}(z)-Y^{eq}_{N_i}(z)\right) \nonumber \\
\frac{dY_{\Delta_\alpha}(z)}{dz}&=&-\sum_{i=1,2}\varepsilon_{i\alpha} K_i
(D_i(z)+S_i(z))\left(Y_{N_i}(z)-Y^{eq}_{N_i}(z)\right)\nonumber \\
&&+\sum_{i=1,2} K_{i\alpha}\sum_\beta W_i(z)\left(A_{\alpha\beta}Y_{\Delta_\beta}(z)
+C_\beta Y_{\Delta_\beta}\right). \label{eq:becomp}
\end{eqnarray}
where $\Delta_\alpha\equiv B/3-L_\alpha$, and $Y_{\Delta_\alpha}$ stands for
the total $\Delta_\alpha$ asymmetry stored in the fermionic flavours, and $z=M_1/T$. 
The washout parameter for various flavors is
\begin{equation}
K_{i\alpha}=\frac{\Gamma(N_i\rightarrow l_\alpha H^*)
+\Gamma(N_i\rightarrow \bar{l}_\alpha H)}{H(M_{N_i})}\label{eq:washout}
\end{equation}
such that $K_i=\sum_\alpha K_{i\alpha}$.
In eq.(\ref{eq:becomp})
the equilibrium number density \cite{Davidson:2008bu, Abada:2008gs} is defined 
as
\begin{equation}
Y^{eq}_{N_i}=\frac{135\zeta(3)}{8\pi^4g_*}R_i^2z^2{\cal K}_2(R_iz)
\xrightarrow []{T>>M_{i}}\frac{135\zeta(3)}{4\pi^4g_*},
\end{equation}
where $R_i=M_i/M_1$. The out-of-equilibrium condition for $N_i$ decay, 
$\Gamma_{N_i} < H(T=M_{N_i})$, requires the lightest right-handed neutrino to acquire mass
$M_{N_1}\gtrsim 4\times 10^{8}$\,GeV \cite{Akhmedov:2003dg}  where 
$H \simeq 1.66 g_*M_{N_k}^2/(M_{\rm Pl}z^2_k)$ is the Hubble 
expansion rate. The thermally averaged decay rates are
$D_i(z)=R_i^2z{{\cal K}_1(R_iz)}/{{\cal K}_2(R_iz)}$
where ${\cal K}_1$ and ${\cal K}_2$ are the first and the second order modified Bessel
functions \cite{Buchmuller:2004nz,Davidson:2008bu}, respectively.
The scattering terms $S_i(z)$ account for Higgs-mediated $\Delta L=1$ scatterings
involving top quark and anti-quark as $S_i(z)=2S_s^i(z)+4S_t^i(z)$  \cite{Buchmuller:2004nz}.
%Or can be written as \cite{Buchmuller:2004nz}
%\begin{equation}
%S_i(z)=\frac{K_S}{6}\left(f^i_{\phi,s}(z)+2f^i_{\phi,t}(z)\right),
%\end{equation}
%where
%\begin{equation}
%f^i_{\phi,t(s)}(z)=\frac{\int_{R_iz}^\infty d\psi 
%f_{\phi,t(s)}\left(\psi/R_i^2z^2\right)\sqrt{\psi}
%{\cal K}_1(\sqrt{\psi})}{R_i^2z^2{\cal K}_2(R_iz)}
%\end{equation}
%and
%\begin{eqnarray}
%f_{\phi,s}(x)&=&\left(\frac{x-1}{x}\right)^2,~~K_S=\frac{9m_t^2}{4\pi^2 g_N v^2_{EW}},\\
%f_{\phi,t}(x)&=&\frac{x-1}{x}\left[\frac{x-2+2a_h}{x-1+a_h}
%+\frac{1-2a_h}{x-1}\ln\left(\frac{x-1+a_h}{a_h}\right)\right]
%\end{eqnarray}
%with $a_h\sim 0.4/R_iz$, we have assumed for thermal Higgs mass $M_h=0.4 T$.
The washout term is $W_i(z)=W^{ID}_i(z)+W_i^S(z)$ where the inverse decay 
contribution is
\begin{equation}
W_i^{ID}(z)=\frac{1}{4}R_i^4z^3{\cal K}_1(R_i z).
\end{equation}
The unit lepton number changing $\Delta L=1$ scattering contributing to washout
is
\begin{equation}
W_i^S(z)=\frac{W_i^{ID}(z)}{D_i(z)}\left(2S_s^i(z)
\frac{Y_{N_i}(z)}{Y^{eq}_{N_i}}+8S^i_t(z)\right).
\end{equation}
%The Bessel functions for all values of $z$ are written as \cite{Buchmuller:2004nz}
%\begin{equation}
%{\cal K}_1(z)\simeq \frac{1}{z}\sqrt{1+\frac{\pi}{2}z}e^{-z},~~
%{\cal K}_2(z)\simeq \frac{1}{z}\left(\frac{15}{8}+z\right){\cal K}_1(z)
%\end{equation} 
The $\Delta L=1$ scattering and related washout from Higgs and lepton mediated
inelastic scattering involving top quark are included in the evolution of
asymmetry  \cite{Buchmuller:2004nz}.

We have ignored the off-shell part of $\Delta L=2$ process in the washout term which
is a good approximation as long as $M_{N_i}/10^{13}<< K_{i}$\cite{Abada:2006ea}. We have also
omitted the $\Delta L=0$ scattering such as $N_iN_j\rightarrow l\bar{l}$,
$N_iN_j\rightarrow HH^*$, $N_il\rightarrow N_jl$, $N_i\bar{l}\rightarrow N_j\bar{l}$
which do not contribute to the washout but can affect the abundance of heavy neutrinos.
When flavor effects are taken into account, they also tend to redistribute the 
lepton asymmetry among flavors. These effects are of higher order in the 
neutrino Yukawa couplings and are expected to have little impact on the final
baryon asymmetry. 
We further neglected the scalar triplet related washout processes,
gauge scatterings, spectator processes, and the higher order processes like
{\small $1\rightarrow 3$} and {\small $2\rightarrow 3$}. The heavy gauge 
bosons processes such as $N_i\, e_R \rightarrow \bar q_R\, q'_R$ and 
$N_i N_i \rightarrow f \bar f$ tend to keep the heavy neutrinos in thermal 
equilibrium, thus reducing the generated lepton asymmetry. This effect 
is practically negligible because RH$\nu$s are much lighter than the RH
gauge bosons. We also ignore such flavour effects
\cite{PSB-Pilaftsis:2014} which are relevant for resonant leptogenesis.  

\subsection{Baryon Asymmetry in the Compact Scenario}\label{sub:n1}
In this scenario the tau lepton flavour state decouples while
the electron and muon states are still coupled.
Thus, a flavour dependent analysis is necessary.
In the two flavour case $Y_{\Delta_{e+\mu}}\equiv Y_{\Delta_e}+ Y_{\Delta_\mu}$, 
$\varepsilon_{i,e+\mu}=\varepsilon_{ie}+\varepsilon_{i\mu}$, 
$K_{i,e+\mu}=K_{ie}+K_{i\mu}$, and the flavour coupling matrices are 
\cite{Antusch:2010ms}
\begin{equation}
A =\begin{pmatrix}
-417/589 & 120/589  \\ 
 30/589 &  -390/589 
\end{pmatrix}\,,~~
C =\begin{pmatrix}
-164/589, & -224/589  \\ 
\end{pmatrix}.\label{eq:A2}
\end{equation}
In this case the baryon asymmetry is expressed as 
\begin{equation}
Y_{\Delta B}=\frac{12}{37}\sum_\alpha Y_{\Delta_\alpha},~~(\rm SM)
\end{equation}
where the factor $12/37$ is due to partial conversion of $\Delta_\alpha$
asymmetry in to baryon asymmetry by non-perturbative sphaleron process
\cite{Khlebnikov:1988sr,Harvey:1990qw}.
The results of BBN \cite{Iocco:2008va} and  PLANCK \cite{Planck15} experiments are
\bea
Y^{BBN}_{\Delta B}&=&(8.10\pm 0.85)\times 10^{-11},\nonumber\\ 
Y^{Planck}_{\Delta B}&=&(8.58\pm 0.22)\times 10^{-11}.\label{BAUexpt}
\eea
Compared to these somewhat higher value of BAU obtained from WMAP 7 years'
data has been reported in ref.\cite{Larson:2010gs}.
% with $Y^{WMAP}_{\Delta B}&=&(8.58\pm 0.22)\times 10^{-11}$
%\begin{figure}[h!]
%\begin{center}
%\includegraphics[scale=0.8]{l1.pdf}
%\includegraphics[scale=0.8]{ol1.pdf}
%\end{center}
%\caption{The baryon asymmetry in $e+\mu$ flavor (dotted blue curve) and
%$\tau$ flavor (dashed red curve). The quartic coupling $\lambda=0.1$.}
%\end{figure}

\begin{figure}[h!]
\begin{center}
\includegraphics[scale=0.4]{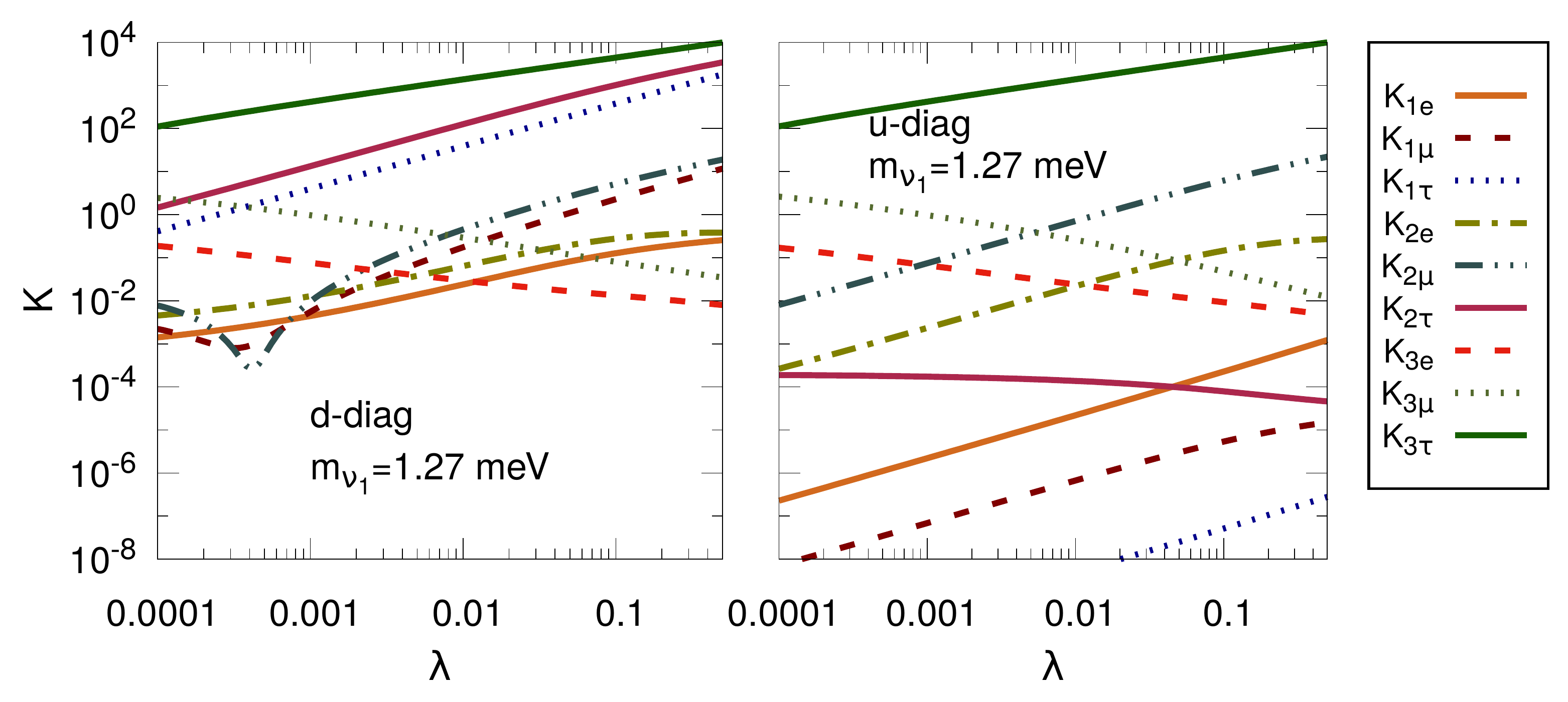}
\end{center}
\caption{Washout factor vs. quartic coupling in the compact spectrum scenario. 
Left(right) panel corresponds to $d(u)$-quark diagonal basis. 
The lightest neutrino mass is kept at 
$m_{\nu_1}=0.00127$\,eV. Other parameters are kept fixed as described in the text.}
\label{fig:kvslambc}
\end{figure}

\begin{figure}[h!]
\begin{center}
\includegraphics[scale=0.4]{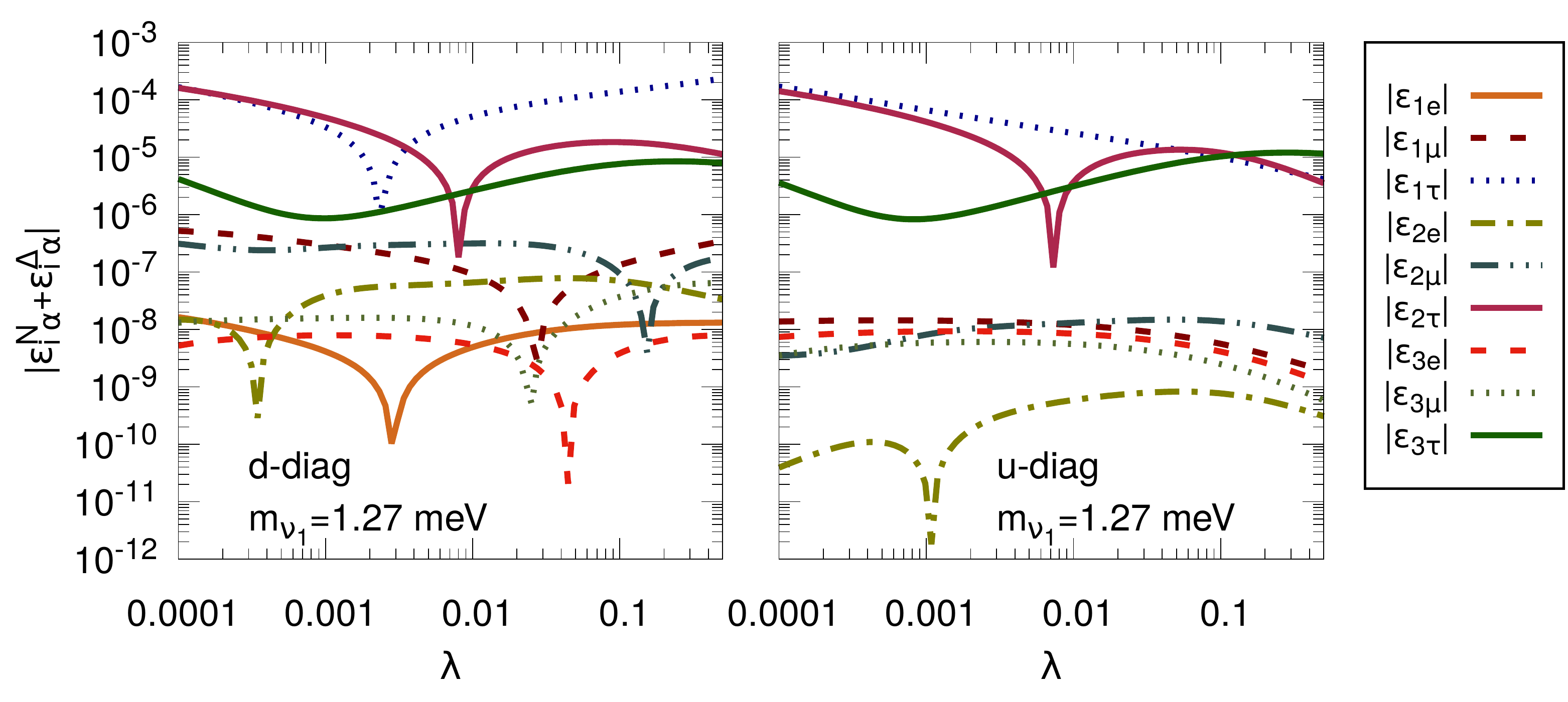}
\end{center}
\caption{CP-asymmetry vs. the quartic coupling in compact spectrum scenario. 
Left (right) panel corresponds to $d (u)$-quark diagonal  basis. The lightest neutrino mass is kept at 
$m_{\nu_1}=0.00127$\,eV. Other parameters are kept fixed as described in the text.}
\label{fig:cpvslambc}
\end{figure}

%\begin{figure}[h!]
%\begin{center}
%\includegraphics[scale=0.8]{l05.pdf}
%\includegraphics[scale=0.8]{ol05.pdf}
%\caption{The baryon asymmetry in $e+\mu$ flavor (dotted blue curve) and
%$\tau$ flavor (dashed red curve). The quartic coupling $\lambda=0.05$.}
%\label{fig:comp}
%\end{center}
%\end{figure}

%\begin{figure}
%\begin{center}
%\includegraphics[height=6cm, width=12cm]{YB.pdf}
%\includegraphics[scale=0.8]{ol1.pdf}
%\caption{Total baryon asymmetry for varying quartic couplings.}
%\label{fig:YBcomp}
%\end{center}
%\end{figure}

%\begin{figure}
%\begin{center}
%\includegraphics[scale=1.0]{ni.pdf}
%\end{center}
%\caption{Right handed neutrino abundance. Due to large 
%washout factor thermal abundance overlaps.}
%\end{figure}

The washout coefficients $K_{i\alpha}$ in the compact spectrum scenario of
RH neutrino masses for the lightest neutrino mass ${m_\nu}_1=0.00127$\,eV  and 
$\lambda\in [0.0001,0.5]$ are plotted in Fig.\ref{fig:kvslambc}.
We see that there are two to four orders of variation in the washout for
the above allowed range of $\lambda$ in both the d-diagonal (left panel) 
and the u-diagonal (right panel) cases. We list the washout parameters
for $\lambda=0.1$ in the case of the d-quark diagonal basis 
\begin{equation}
K=\begin{pmatrix}
1.27\times 10^{-1} &  2.28 &  3.81\times 10^{2} \\
   2.77\times 10^{-1} &  5.16 &  1.03\times 10^{3} \\
    1.34\times 10^{-2} &  8.04\times 10^{-2} &  4.44\times 10^{3} \\
\end{pmatrix}.
\end{equation}
In the $u$-quark diagonal basis the washout parameters are
\begin{equation}
K=\begin{pmatrix}
 2.27\times 10^{-4} &  5.37\times 10^{-6} &  5.14\times 10^{-8} \\
   1.46\times 10^{-1} &  6.19 &  7.88\times 10^{-5} \\
   9.23\times 10^{-3}  & 4.75\times 10^{-2} &  4.45\times 10^{3} \\
\end{pmatrix}.
\end{equation}
Our observations in the two cases are summarized below.\\

\par\noindent{\bf (a)The $d$-quark diagonal basis:}\\
 We note that $K_i=\sum_\alpha K_{i\alpha}\sim (300-4000)$. 
Therefore the system is in strong washout regime for most of the parameter space. 
The asymmetry is determined by a balance between production and destruction. The final asymmetry
freeze occurs at the decoupling of washout with $z_f\sim (7-10)$. In the single flavour
analysis the lepton asymmetry is approximated as \cite{Fong:2013wr}
\begin{equation}
Y_{\Delta L}(\infty)\simeq \frac{\pi^2}{6z_fK_1}\varepsilon_1Y_{N_1}^{\rm eq}(0).
\end{equation}
Using the values of $K_i$ from Fig.~\ref{fig:kvslambc} and 
$\varepsilon_1=\sum_\alpha \varepsilon_{1\alpha}$ from Fig.~\ref{fig:cpvslambc} 
we can easily achieve the required lepton asymmetry. In fact it may lead to
a constraint on quartic coupling $\lambda$.\\ 
%Increasing $\lambda$ will increase $K$ as shown in (Fig.\,\ref{fig:kvslambc}) 
%while CP-asymmetry either decreases or stays of same order 
%(Fig.\,\ref{fig:cpvslambc}), therefore effectively decreasing 
%total asymmetry $Y_\Delta$.  

\par\noindent{\bf (b) The $u$-diagonal basis:}\\ 
We note that, since $K_1=\sum_\alpha K_{1\alpha}<<1$, this is  
 a very weak washout regime. Ignoring thermal effect on CP-asymmetry 
and assuming zero initial abundance in the weak washout regime with initial
thermal  
abundance $Y_{N_1}(z=0)=Y^{\rm eq}_{N_1}(z=0)$ \cite{Fong:2013wr} gives
\begin{equation}
Y_{\Delta}(\infty)\simeq \varepsilon_1 Y_{N_1}^{\rm eq}(0).
\end{equation}
If there is already an initial amount of asymmetry left 
over, say through $N_2$ decay, it will not be washed out because the system is in weak washout regime.
But with zero initial abundance, $Y_{N_1}(z=0)=0$ \cite{Fong:2013wr}
\begin{equation}
Y_{\Delta L}(\infty)\simeq \frac{27}{16}\varepsilon_1 K_1^2Y_{N_1}^{\rm eq}(0).
\end{equation}
We note that even if we assume initial thermal abundance $Y_{N_1}^{\rm eq}(0)\sim 0.0039$,
the CP-asymmetry $\varepsilon_1 \sim 10^{-4}-3\times 10^{-6}$ (Fig.~\ref{fig:cpvslambc}) and 
$K\sim 10^{-7}-10^{-3}$ (Fig.~\ref{fig:kvslambc}). Therefore the generated asymmetry
would be determined by initial abundance and, in the zero initial abundance scenario, the required
lepton asymmetry can not be produced for any parameter
value. Therefore the flavour independent analysis
in the $u$-quark diagonal scenario with zero initial abundance of $Y_{N_1}$ fails to give the required 
asymmetry. 
%Thus the left over asymmetry generated during $N_2$ decay survives. Since $N_2$ is just one 
%order heavier and $K_2\sim 6$, even an initial zero abundance of $Y_{N_2}$ will reach equilibrium
%around $z_2 (\equiv M_2z/M_1)=1$, and since $\varepsilon_{2}\sim 10^{-6}-10^{-7}$, will leave 
%some initial asymmetry which will survive in low washout of $N_1$. 

\begin{figure}[h!]
\begin{center}
\includegraphics[scale=0.4]{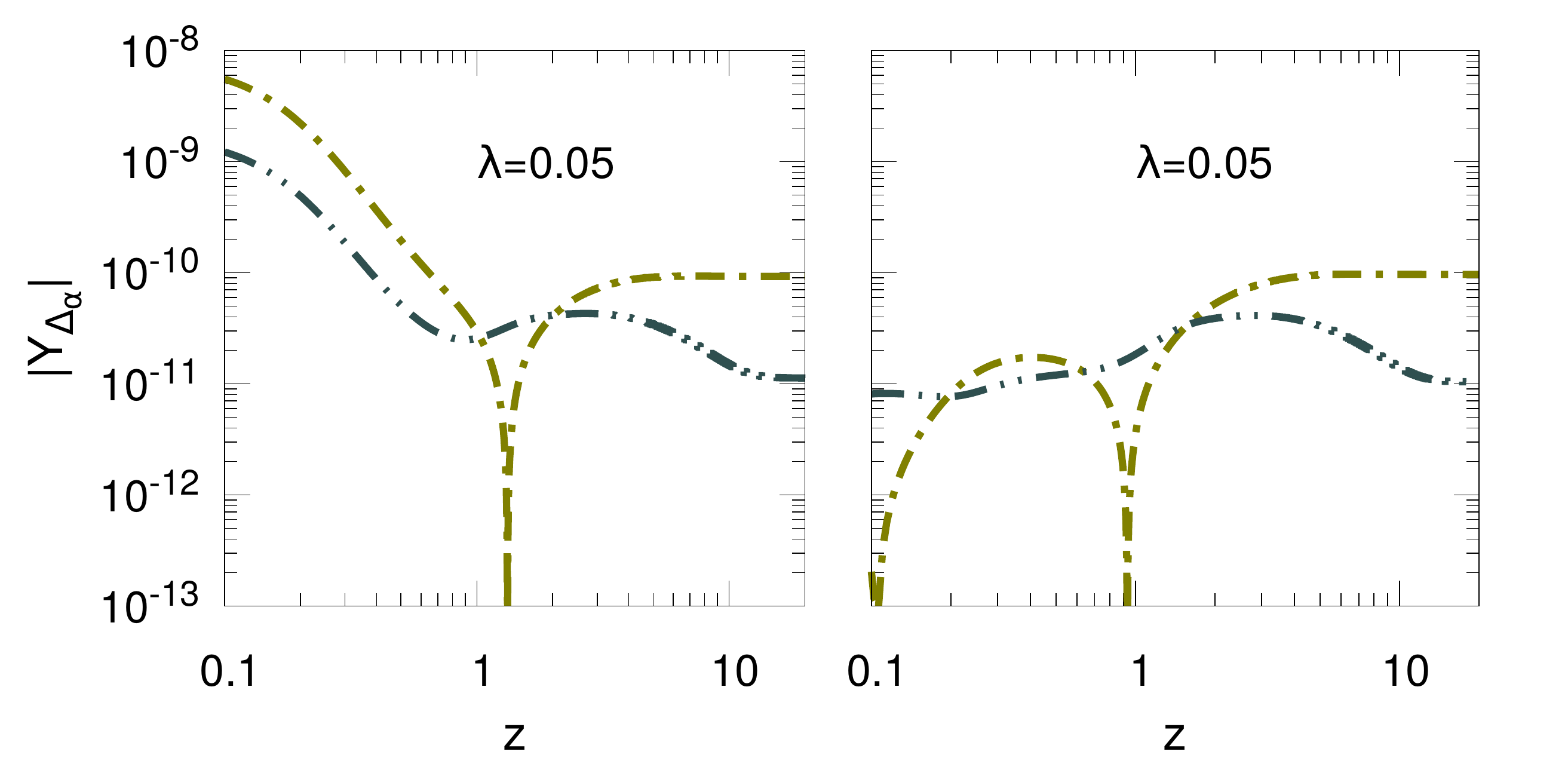}
\caption{The baryon asymmetry in $e+\mu$ flavours (double-dot-dashed blue curve) and
$\tau$ flavor (dot-dashed  curve) for the $u$-quark diagonal basis and
  compact spectrum RH$\nu$ mass scenario. 
Left (right) panel correspond to non-zero (zero) initial thermal abundance. 
The quartic coupling $\lambda=0.05$.}
\label{fig:comp}
\end{center}
\end{figure}

On the other hand a flavor dependent analysis can enhance the asymmetry. The flavour
dependent lepton asymmetry is analyzed using Boltzmann equations (\ref{eq:becomp}) and 
is shown in Fig.~\ref{fig:comp} for $u$-quark diagonal basis. Thus in flavoured 
analysis we find that final lepton asymmetry is independent of initial abundance and
is close to the experimental value for $\lambda <0.05$. This explicitly shows that $N_2$
decay contributes to lepton asymmetry which is not completely washed out in the $N_1$ decay.

\begin{figure}[h!]
\begin{center}
\includegraphics[height=12cm, width=14cm]{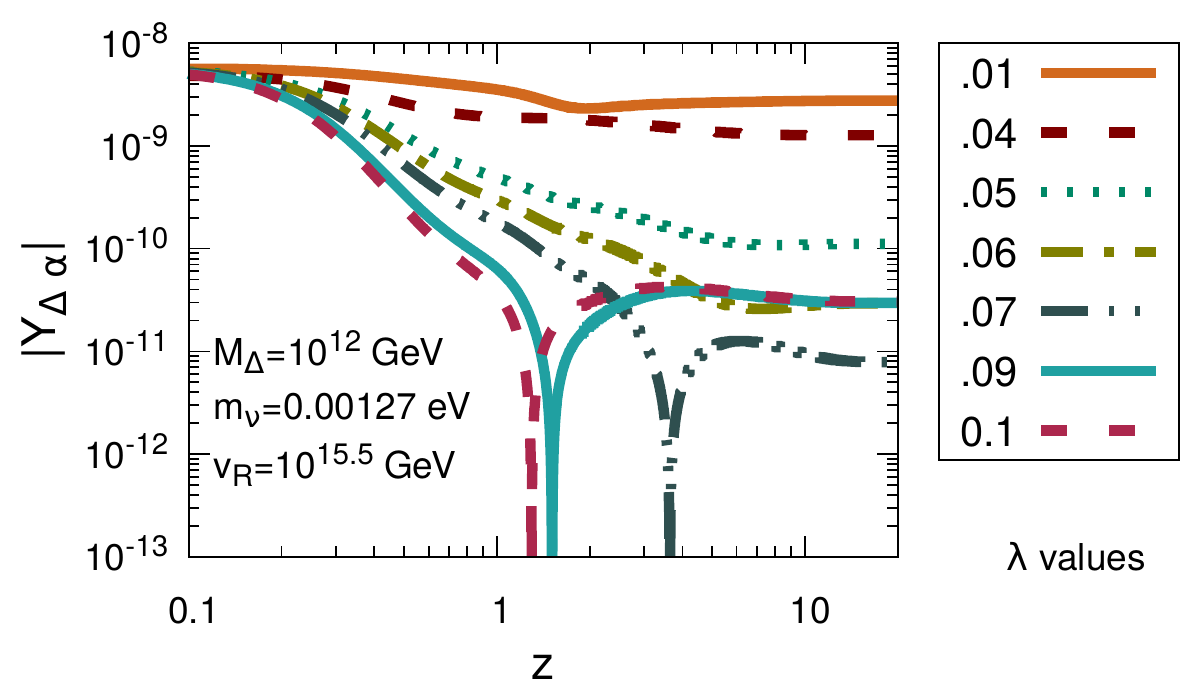}
\caption{Total baryon asymmetry vs. $z$ for different values of the
  quartic coupling  in  the compact spectrum scenario with Dirac neutrino mass
  matrix determined in the
  $d$-quark diagonal basis as described in the text.}
\label{fig:YBcomp}
\end{center}
\end{figure}

The reason for doing flavoured analysis is that there are enhancements in the final
asymmetry compared to the unflavoured case. Using $d$-quark diagonal 
basis Fig.\,\ref{fig:YBcomp} shows the variation of 
total asymmetry with respect to quartic coupling  for
a fixed value of the scalar triplet mass $M_\Delta=10^{12}$\,GeV, 
$v_R=10^{15.5}$\,GeV, and the lightest neutrino mass
$m_{\nu_1}=0.00127$\,eV in normalyy ordered case. Similar is the effect in the $u$-quark diagonal basis.

%Baryon asymmetry $N_k$-decay $N_j$ mediated
%$\eta_B=(3.0069\times 10^{-10}, 2.4137\times 10^{-14},-1.2186\times 10^{-15})$.
%Baryon asymmetry $N$-decay $\Delta_L$ Mediated
%$\eta_B=(3.3737\times 10^{-10}, -2.1161\times 10^{-14}, -1.6950\times 10^{-13})$.
%Changing the light neutrino mass will change the matrix $f$ and hence the
%mass of right-handed neutrinos. 
%We have therefore plotted the produced 
%baryon asymmetry versus lightest neutrino mass in fig.\,\ref{fig:fixm} for 
%fixed $M_\Delta=10^{12}$\,GeV, and in fig.\,\ref{fig:fixl} for fixed
%quartic coupling $\lambda=0.1$.

%\begin{figure}
%\begin{center}
%\includegraphics[scale=1.0]{fixM.pdf}
%\end{center}
%\caption{Baryon asymmetry $N_1$ decay, in normal hierarchy of light neutrinos
%for fixed $M_\Delta=10^{12}$\,GeV. Solid curves correspond to asymmetry 
%generated due to right handed neutrino mediating the radiative correction, 
%and dashed curves due to triplet scalar mediation.}\label{fig:fixm}
%\end{figure}
%
%\begin{figure}
%\begin{center}
%\includegraphics[scale=1.0]{fixL.pdf}
%\end{center}
%\caption{Baryon asymmetry $N_1$ decay, in normal hierarchy of light neutrinos for
%fixed $\lambda=0.1$. Solid curves correspond to asymmetry generated due to right
%handed neutrino mediating the radiative correction, and dashed curves due to
%triplet scalar mediation.}\label{fig:fixl}
%\end{figure}

\subsection{Baryon Asymmetry in the Hierarchical Scenario}\label{sub:n2}
The Davidson-Ibarra bound is not respected in the hierarchical
spectrum scenario 
of RH$\nu$ (see Fig. \ref{fig:hcmass}). In such a case there is the possibility of
leptogenesis if asymmetry is produced by the decay of $N_2$. Lower 
bound on the lightest RH$\nu$ is passed to $M_{N_2}\gtrsim 10^{10}$\,GeV.
The $N_2$-dominated leptogenesis can be successful if there is a heavy 
neutrino, or triplet scalar with $M_{N_3}, M_{\Delta_L}> M_{N_2}$, and the 
washout from the lightest RH$\nu\,(N_1)$ is circumvented. Since 
$M_{N_1}<<10^9$\,GeV the lepton flavour states become incoherent and 
the washout acts separately on each flavour asymmetry. We need 
to solve Boltzmann equations at the production phase 
with $z_2=M_2/T$, and at the washout phase with $z_1=M_1/T$ \cite{Antusch:2010ms}.
We note from the Fig.\,\ref{fig:cpah} that the CP-asymmetry due to
$N_1$ decay $\varepsilon_i=\sum_\alpha \varepsilon_{i\alpha}$ is very 
small compared to CP-asymmetry due to $N_{2,3}$ decays. The decay
and washout are also suppressed by a factor $M^2_1/M^2_3 (\sim 10^{-14}-10^{-15})$ and 
$M_1^2/M_2^2 (\sim 10^{-9}-10^{-10})$. Also we note that
in the scenario $M_3\gtrsim 10^{12}$\,GeV$>>M_2>10^9$\,GeV$>>M_1$, the
role of $N_3$
becomes indistinct by the time asymmetry is produced due to $N_2$ decay and when
washout is active. Thus $N_{1,3}$ do not contribute to asymmetry
generation
at the $N_2$ decay phase and  we can write
\begin{eqnarray}
\frac{dY_{N_2}(z_2)}{dz_2}&=&-K_2(D_2(z_2)+S_2(z_2))\left(Y_{N_2}(z_2)-Y^{eq}_{N_2}(z_2)\right)
\label{eq:n2be}\\
\frac{dY_{\Delta_\alpha}(z_2)}{dz_2}&=&-\varepsilon_{2\alpha} K_2
(D_2(z_2)+S_2(z_2))\left(Y_{N_2}(z_2)-Y^{eq}_{N_2}(z_2)\right)\nonumber \\
&&+K_{2\alpha}\sum_\beta W_2(z_2)\left(A_{\alpha\beta}Y_{\Delta_\beta}(z_2)
+C_\beta Y_{\Delta_\beta}(z_2)\right).\label{eq:deln2be}
\end{eqnarray}

\begin{figure}[h!]
\begin{center}
\includegraphics[scale=0.4]{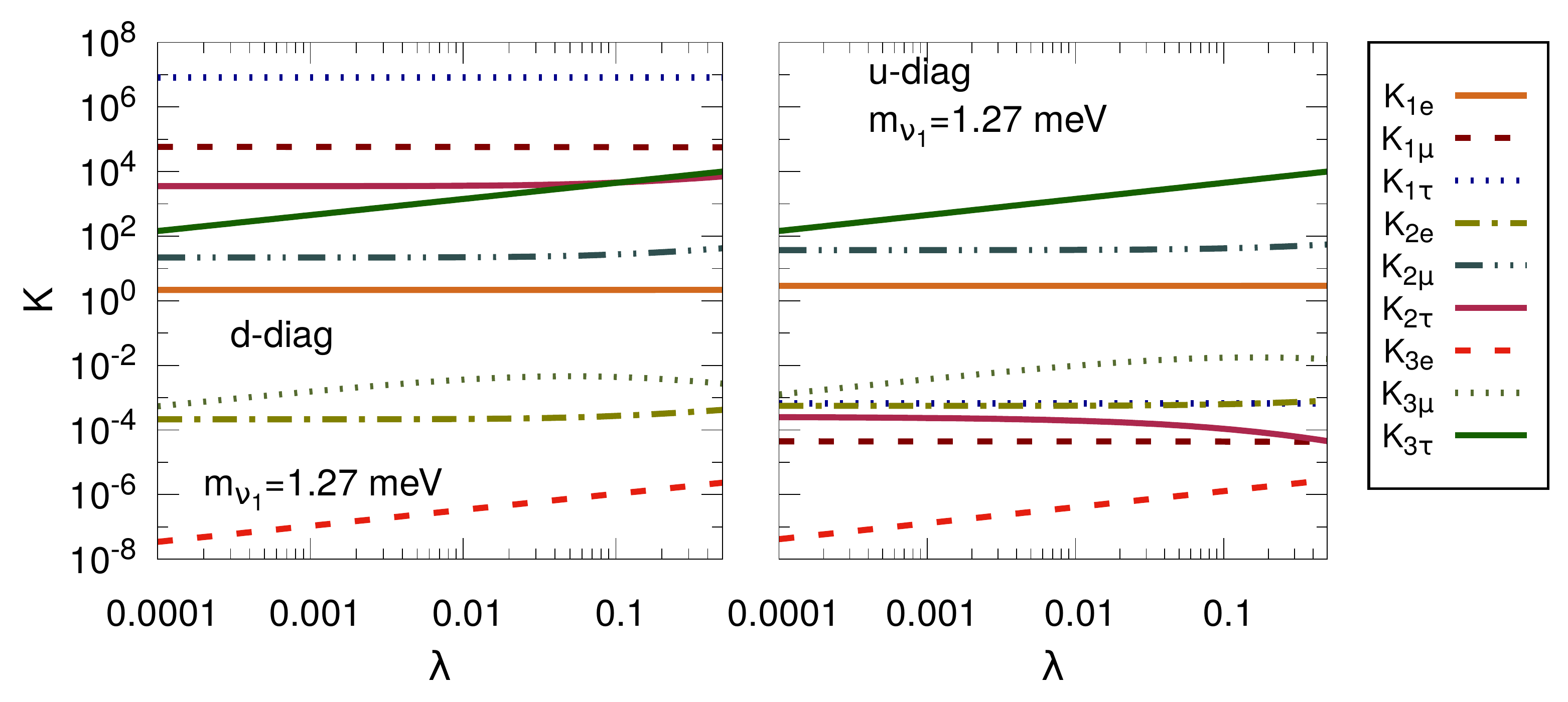}
\end{center}
\caption{Washout factor vs. quartic coupling in the hierarchical
  spectrum scenario of 
RH$\nu$. The 
left (right) panel corresponds to the $d(u)$-quark diagonal basis. 
The lightest neutrino mass is kept at 
$m_{\nu_1}=0.00127$\,eV. Other parameters are kept fixed as described in the text.}
\label{fig:hkvslambh}
\end{figure}

\begin{figure}[h!]
\begin{center}
\includegraphics[scale=0.4]{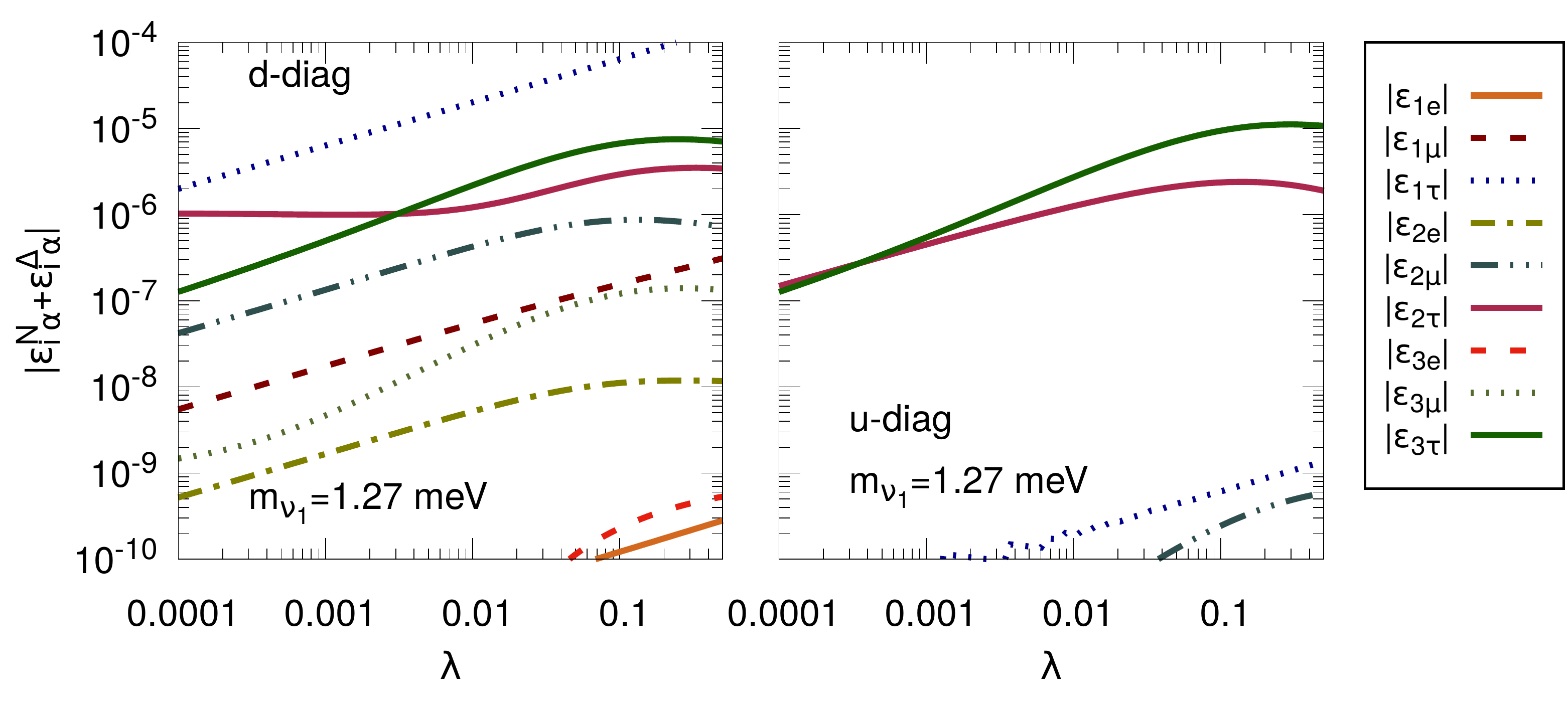}
\end{center}
\caption{CP-asymmetry vs. quartic coupling in the hierarchical
  spectrum scenario
of RH$\nu$. The  
left (right) panel corresponds to the $d(u)$-quark diagonal basis. 
The lightest neutrino mass is kept at 
$m_{\nu_1}=0.00127$\,eV. Other parameters are kept fixed as described in the text.}
\label{fig:cpvslambh}
\end{figure}

The flavour coupling matrices in the production phase are the same as given
in eq.\,(\ref{eq:A2}). For $T\lesssim 10^9$\,GeV, the muon Yukawa interaction
also gets equilibrated. Then the flavour coupling matrices are 
\cite{Antusch:2006cw,Antusch:2010ms}
\begin{equation}
A\ =\begin{pmatrix}
-151/179 & 20/179 &  20/179 \\ 
 25/358 &  -344/537 &  14/537 \\ 
 25/358 & 14/537 & -344/537
\end{pmatrix}\,,~~C=-(37/179, 52/179, 52/179).\label{eq:A3}
\end{equation}
The washout parameters in the $d$-quark diagonal basis for ${m_\nu}_1=0.00127$\,eV
and $\lambda=0.1$ are
\begin{equation}
K=\begin{pmatrix}2.157 & 58072 & 8.19\times 10^6 \\
0.00021 & 21.80 & 3545.8 \\
1.1\times 10^{-7} & 0.00154 & 450.1 
\end{pmatrix}.
\end{equation}
 In the $u$-quark diagonal basis they are
\begin{equation}
K=\begin{pmatrix}
2.899 &  4.42\times 10^{-5} &  6.64\times 10^{-4} \\
5.57\times 10^{-4}  & 37.11 &  2.346\times 10^{-4} \\
1.297\times 10^{-7}  & 0.0037 &  451.343 \\
\end{pmatrix}.
\end{equation}
The washout factors and the CP-asymmetries for different flavours as
a function of quartic coupling are shown in Fig.\ref{fig:hkvslambh} 
and in Fig.\ref{fig:cpvslambh}, respectively, for the
$d$-quark diagonal (left panel) and the $u$-quark diagonal (right panel) 
bases in each case. Notice that in the $d$-quark diagonal basis
$K_{1\alpha}>>1$ for $\alpha=\mu,\tau$.
Therefore any such type of flavoured asymmetry produced during $N_2$ decay will be washed out during 
the $N_1$ decay. But since $K_{1e} \simeq 2$ the corresponding
flavoured asymmetry will be washed out only partially. However, in the $u$-quark diagonal basis, $K_{1\alpha}(\alpha \neq e)<<1$.
Therefore the corresponding flavour asymmetries produced during $N_2$ decay
would survive. Also noting that in this basis $K_{1e} \sim 2.8$, the $e$-asymmetry generated by the $N_2$ decay will be only partially washed out by the $N_1$ 
decay. Also, noting  
from Fig.\ref{fig:cpvslambh} that $\varepsilon_{2\tau}$ is significantly large 
, it may produce the required amount of asymmetry. The complete flavoured
analysis scenario is discussed below.\\

%In the hierarchical scenario in d-quark diagonal basis,
%$K_{1e}\sim 1$ while $\varepsilon_{2e}\sim 10^{-8}$, while
%in the u-quark diagonal basis $K_{1\alpha}\lesssim 1$ leading to the largest 
%asymmetry $\varepsilon_{2\tau}$. 
%Variation in $CP$-asymmetry with the change in lightest
%active neutrino mass in this scenario has been already shown in 
%Fig.\,\ref{fig:cpah}.
%The phantom terms appearing in this scenario would also 
%contribute if the initial asymmetry is assumed. 

\begin{figure}[h!]
\begin{center}
\includegraphics[scale=0.4]{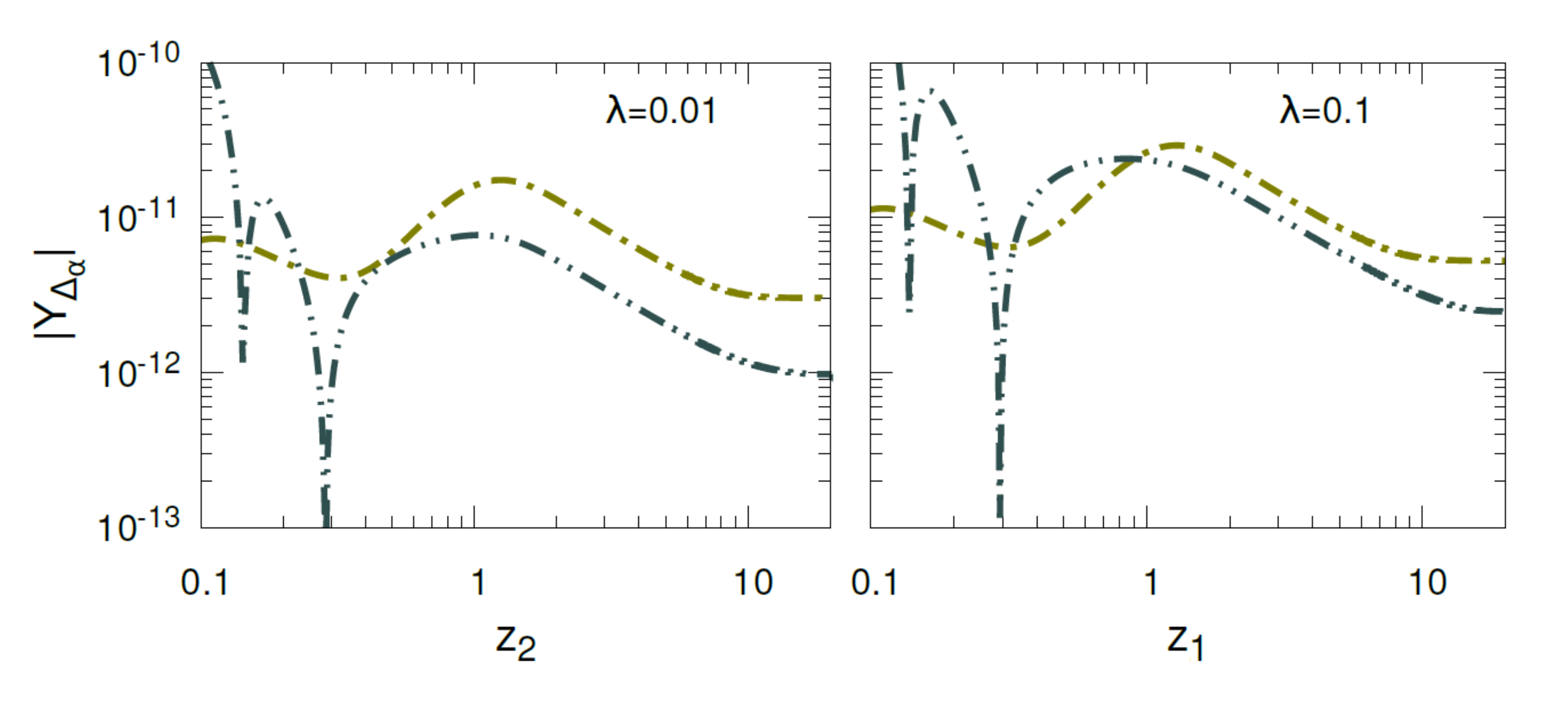}
\end{center}
\caption{The asymmetry with $e+\mu$ flavors (double-dot-dashed blue curve)
  and the
$\tau$ flavor (dot-dashed curve) due to $N_2$ decay. The left (right) panel
represents our estimations for  quartic 
coupling $\lambda=0.01(0.1)$.}
\label{fig:flavn1d}
\end{figure}

With the washout caused due to the $N_1$ decay, the solutions to Boltzmann equations can be achieved
by the substitution $2\rightarrow 1$ everywhere. Since $N_{2,3}$ abundance has
vanished below $10^9$\,GeV, the corresponding equations are redundant. 
We also note from Fig.\,\ref{fig:cpah} that the CP-asymmetries 
$\varepsilon_{i\alpha}$ are negligibly small, therefore the first term in the 
RHS of corresponding equation in eq.\,(\ref{eq:deln2be}) in the $N_1$ 
decay can be ignored when $K_1$ is not very large. 
%In the d-quark diagonal basis
%$K_1$ is very large, therefore the $Y_{N_1}$ reaches equilibrium very quickly
%and never leaves it thereafter. 
This results in the redundancy of the equation for $N_1$ in 
 eq.\,(\ref{eq:n2be}) and we need to solve
only
\begin{eqnarray}
\frac{dY_{\Delta_\alpha}(z_1)}{dz_1}&=& K_{1\alpha}\sum_\beta W_1(z_1)
\left(A_{\alpha\beta}Y_{\Delta_\beta}(z_1)
+C_\beta Y_{\Delta_\beta}(z_1)\right).\label{eq:deln1be}
\end{eqnarray}
\begin{figure}[h!]
\begin{center}
\includegraphics[scale=0.4]{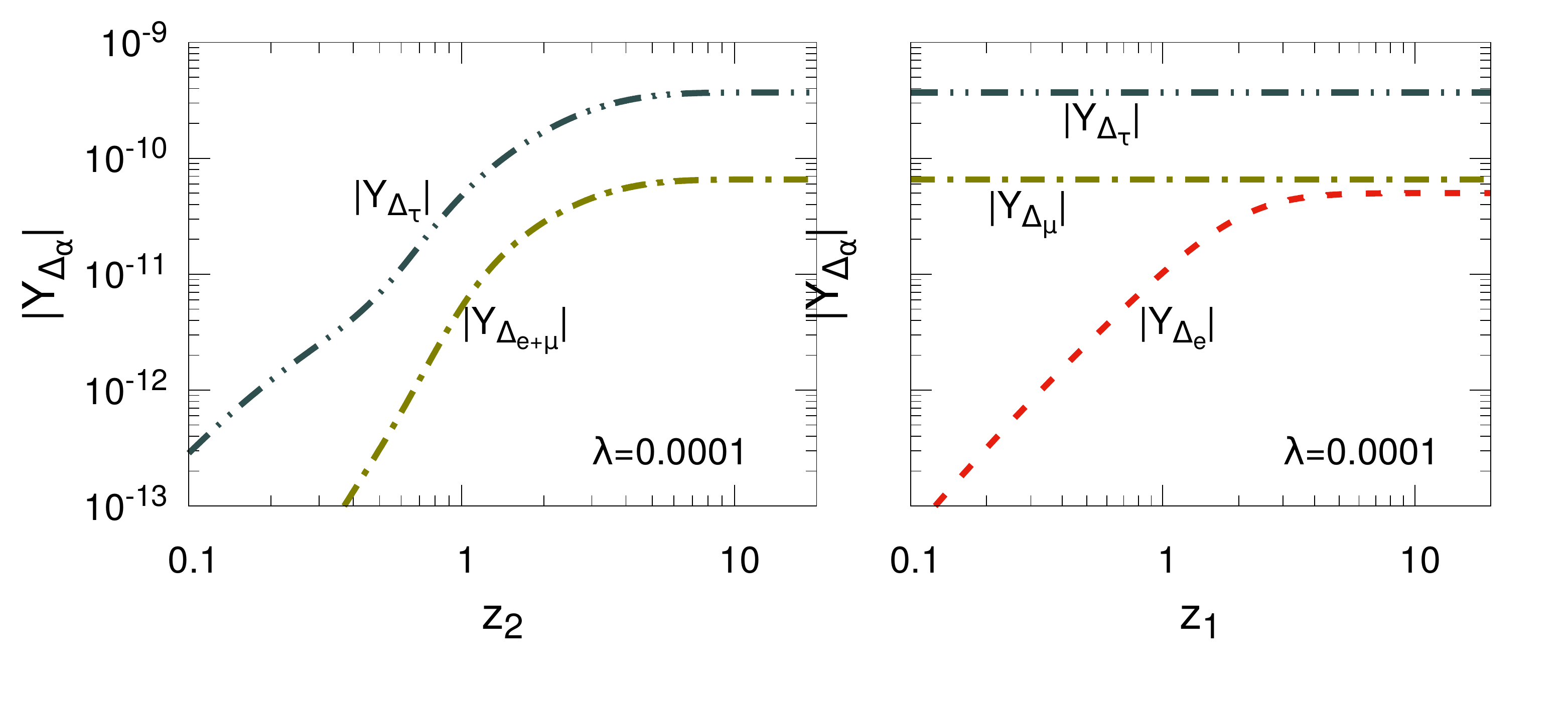}
\end{center}
\caption{The flavor asymmetries in $e+\mu$ and $\tau$ flavors (left panel) and
separately for $e$,$\mu$ and $\tau$ flavors (right panel). The quartic 
coupling has been fixed at $\lambda=0.0001$.}
\label{fig:flavn1u}
\end{figure}
The washout from the lightest RH$\nu$ is more efficient which acts on 
the whole of the generated asymmetry.
%Since $K_{1\tau}>>1$, the $Y_{\Delta_\tau}$ is washed out immediately.
%The other asymmetries are washed out by more than an order of the survived
%asymmetry at the $N_2$ decay. 
We found that in the $d$-quark diagonal basis, 
the asymmetry $Y_{\Delta_\alpha}$ produced by the $N_2$ decay as shown in Fig.\ref{fig:flavn1d} 
is itself much smaller than the experimentally observed asymmetry. There is no
way to enhance it at the stage of $N_1$ decay  in the case of $d$-quark diagonal basis leading to 
 insufficient asymmetry. We also note from 
Fig.~\ref{fig:hkvslambh} and  Fig.~\ref{fig:cpvslambh} 
that variation in quartic coupling is not going to help in enhancing the depleted asymmetry.

On the other hand in the $u$-quark diagonal basis $K_{1e}\sim 2$ and $K_{1\mu(\tau)}<<1$, the 
asymmetries may survive the washout during the $N_1$ decay. In Fig.\ref{fig:flavn1u} 
%In Fig.\,\ref{fig:flavn1d} we show that the flavor asymmetry is smaller 
%than the required asymmetry by almost one to two orders for allowed
%values of parameters. Also the asymmetry  does not 
%appear to be approaching the experimental value for the allowed parameters 
% considered here. But for the baryon asymmetry estimated using much smaller values of quartic
%coupling $\lambda=0.0001$,
 we have shown  solutions to Boltzmann equations where the flavour asymmetries are found to reach the experimental value. The left-panel 
of the figure corresponds to asymmetry produced during $N_2$ decay and
the right-panel corresponds to the asymmetries surviving the $N_1$
decay washout. The results have been computed 
for $\lambda=0.0001$ i.e. for the parameters where CP-asymmetry is the
smallest as indicated in the Fig.\ref{fig:cpvslambh}.
As a matter of fact the behaviours of all the three individual asymmetries in the right-panel clearly follow analytically as solutions to  eq.\,(\ref{eq:n2be})
for which the coupling parameters are given in eq.(\ref{eq:A3}). Noting that
$|A_{ee}|\sim 1$ but $|A_{e\mu}|\sim |A_{e\tau}| << 1$, and $C_e\sim C_{\mu} \sim C_{\tau}<< 1$ gives the rising behaviour of $|Y_{\Delta_e}|$ from  eq.\,(\ref{eq:n2be}) as $K_{1e}\sim 2$. But because of  the negligible values of  $K_{1\mu}$ and    
$K_{1\tau}$,  eq.\,(\ref{eq:n2be}) gives constant behaviours for $|Y_{\Delta_\mu}|$
and $|Y_{\Delta_\tau}|$ as shown in the right-panel of Fig.\ref{fig:flavn1u}.

Using type-I seesaw and $N_2$ dominated flavoured leptogenesis it has
been shown that parts of $e$ and $\mu$ asymmetries, designated as
phantom terms \cite{Antusch:2010ms}, can completely escape washouts
due to the lightest RH$\nu$ $N_1$ decay. Such phantom terms can give large
contribution to the asymmetry resulting in a large $B-L$ asymmetry generation by the
$N_1$ wash outs. The $N_2$ dominated leptogenesis generated due to
such terms has been termed as ``phantom''
leptogenesis. In this work \cite{Antusch:2010ms} each of the phantom terms being proportional to the $N_2$
abundance, the phantom terms vanish in the case of zero initial number density of the
heavier RH$\nu$ i,e $N_2$.\\ 

However in a subsequent investigation \cite{Blanchet:2011} phantom terms have been shown
to emerge as a generic feature of flavoured leptogenesis. They have to
be  taken into account even for initially vanishing RH$\nu$
abundances. In the strong washout regime the phantom terms have
been also shown to give a contribution independent of initial
conditions.\\

In the present case with hybrid seesaw as the origin of neutrino masses
and leptogenesis, we find that      
even though we have ignored any such phantom term  in the three flavour analysis, the $N_1$ decay does not wash out the produced asymmetry at all.
Also since $K_{1e}\sim 1$ it helps increasing $Y_{\Delta_e}$ during the second phase of decay.
Thus the conclusion of this analysis is that, in the hierarchical spectrum of 
RH$\nu$s, the production of the
observed baryon asymmetry of the universe in heavy neutrino decays is favoured when Dirac mass matrix 
is such that it is derived from a GUT in the flavour basis satisfying
$Y_u(M_Z)=Y^{\rm diag}_u(M_Z)$.\\    

 To summarize this section, we have attempted to generate the right
 value of BAU through lepton asymmetry produced by the hybrid seesaw
 mechanism where the three heavy RH$\nu$s and a LH triplet scalar
 decay directly or act as mediators in the one-loop
 Feynman diagrams.  Two
 classes of heavy RH$\nu$ spectra are found to be predicted by the neutrino
 oscillation data: compact and hierarchical. We have carried out
 complete flavor dependent analysis in both these cases. We have also
 examined the possibility of basis dependence that
 determines the Dirac neutrino mass matrix at the GUT scale  by
 choosing either the $u$-quark diagonal basis , or  the $d$-quark
 diagonal basis. Rigorous
 solutions to the Boltzmann equations are exploited in every case.
  In the compact spectrum case, the decay of the lightest RH$\nu$ which is heavier than
  the Davidson-Ibarra bound, produces the desired BAU in both the
  choices of the  Dirac
  neutrino mass matrix. This is shown in Fig.8 and Fig.9.  In the
  hierarchical spectrum scenario the lightest RH$\nu$ is much lighter than the  
Davidson-Ibarra bound. The right value of CP-asymmetry is generated predominantly by
the decay of heavier RH$\nu$ $N_2$ that also  survives the wash out
caused by the lightest $N_1$. Successful generation of BAU shown in
Fig.13 is possible with the Dirac neutrino mass matrix determined in
the $u$-quark diagonal basis.  Although direct decay of the LH scalar
triplet itself does not produce the lepton asymmetry to produce the
required BAU, its one loop mediation to the RH$\nu$ decay vertex
correction generates the desired asymmetry which is comparable to
other contributions. Thus the role of the LH triplet predicted by the
matter parity based SO(10) model is emphasized in the generation of
BAU.

\section{Fermionic Triplet as  Dark Matter Candidate}\label{sec:dm}
\subsection{General Considerations with Matter Parity}
Usually the prospective DM candidates  are
 accommodated in model extensions by imposing additional discrete
symmetries for their stability. But as noted in Sec.\ref{sec.1}
an encouraging aspect of non-SUSY $SO(10)$ is that
\cite{Kadastic:dm1,Kadastic:dm2,Hambye:Rev} matter parity is available as an
intrinsic gauged discrete symmetry if the neutral component of the RH
higgs triplet $\Delta_R(1, 3, -2, 1) \subset {126}_H \subset SO(10)$
is assigned GUT scale VEV to break the gauge symmetry leading to the
SM Lagrangian. As the Higgs particle possesses even value of $|B-L|$,
the vacuum with SM gauge symmetry conserves matter parity $P_M=(-1)^{3(B-L)}$  
. This enables to identify the SO(10) representations to be identified
with odd value of $P_M$ for $16$, $144$,$560$,....but with even $P_M$ for
$10$, $45$, $54$, $120$, $126$, $210$, ${210}'$,$660$ ..... Then it
turns out that 
the would-be DM fermions must be in the non-standard fermionic
representations ${10}_F,{45}_F,{54}_F,{120}_F,{126}_F,
{210}_F$.....Thus the smallest representation to provide a doublet
fermion with hypercharge $Y=\pm 1$ is ${10}_F$ and the hyperchargeless
triplet needed for this model building is in the next larger
representation ${45}_F\subset$ SO(10).
 
Similarly if it is desired to construct models with  scalars as DM candidates, they
must belong to the odd $P_M$ scalar representations ${16}_H, {144}_H$.....
Whereas the phenomenology of scalar DM has been emphasized in
\cite{Kadastic:dm1,Kadastic:dm2}, the triplet fermionic DM has
been found suitable in
model construction in \cite{Frig-Ham:2010,psb:2010}. In addition, the color
octet fermions have been found to be essential at high scale
$M_{C_8}\ge 10^{10}$ GeV
\cite{Frig-Ham:2010}. The importance of
various other types of DM along with the triplet fermions of both
types of chiralities has been also discussed in high intermediate scale models
\cite{Mambrini:2015}.

An important advantage of using triplet or doublet fermions over
scalars as DM is that in the limit of zero chiral fermion masses, a $U(1)$
global lepton symmetry of the SM is restored. Thus a value of the
fermion mass substantially lighter than the GUT scale is naturally
protected by this global symmetry in the 't Hooft sense. \cite{Gthooft:1975}.   
On the other hand if a scalar component is used as DM, its mass
lighter than the GUT scale has to be obtained by additional
fine-tuning in the Lagrangian . Also  matter parity conservation
forbids it from acquiring any VEV.

\subsection{Light Non-Standard Fermion Masses from SO(10)}
In this model with the SM gauge symmetry below the GUT scale, a triplet fermionic DM
candidate with zero hypercharge appears to be more appropriate with its mass of the order of TeV scale for gauge
coupling unification as would be shown below in Sec.\ref{sec:uni}.
The neutral component of 
 fermionic triplet $\Sigma_F(1,3,0) \subset {45}_F \subset SO(10)$
 would act as a  cold dark matter candidate. 
For accurate coupling unification we also need a Majorana-Weyl type color octet fermion
$C_8(8,1,0)$ at lower scale.  
Using Yukawa interaction via higher dimensional non-renormalizable operators, the light
triplet fermion mass $\subset {45}_F$ has been obtained in ref.
\cite{Frig-Ham:2010}. But both the lighter values of masses of the triplet
fermion and the octet fermion can
be obtained easily from the renormalizable SO(10) Yukawa Lagrangian at
the GUT scale. In the notation ${45}_F=A_F$, ${54}_H=E$, and
${210}_H={\bf \Phi}$, the relevant GUT scale Lagrangian is 
\begin{eqnarray}
&&-{\cal L}_{Yuk}= A_F\left(m_A+h_p{\bf \Phi}+h_eE
  \right)A_F,\label{Yukgut}
\end{eqnarray}
where $m_A\simeq M_U$ and $h_i (i=p, e)$ are Yukawa couplings. Using GUT scale vacuum
expectation values for the  singlet in E and three singlets in
 ${\bf  \Phi}$ \cite{Fukuyama:2006}, the mass formulas for different
components of ${45}_F$ are
\begin{eqnarray}
&&m(3, 1, 2/3)=m_A+\sqrt{2}h_p\frac{\bf \Phi_2}{3}-2h_e\frac{<E>}{\sqrt
    {15}},\nonumber\\
&&m(3, 2,1/6)=m_A-h_p\frac{\bf \Phi_3}{3}+h_e\frac{<E>}{2\sqrt
    {15}},\nonumber\\
&&m(3,2,-5/6)=m_A -h_p\frac{\bf \Phi_3}{3}+h_e\frac{<E>}{2\sqrt
    {15}},\nonumber\\
&&m(1,1,1)=m_A+\sqrt{2}h_p\frac{\bf \Phi_1}{\sqrt 3}+\sqrt{3}h_e\frac{<E>}{\sqrt
    {5}},\nonumber\\
&&m(1,1,0)=m_A+2\sqrt{2}h_p\frac{\bf \Phi_2}{3}+\sqrt{3/5}h_e<E>,\nonumber\\
&&m'(1,1,0)=m_A+2\sqrt{2}h_p\frac{\bf \Phi_2}{3}-2h_e\frac{<E>}{\sqrt
    {15}},\nonumber\\
&&m_{\rho_8}(8,1,0)=m_A+\sqrt{2}h_p\frac{\bf \Phi_2}{3}-2h_e\frac{<E>}{\sqrt
    {15}},\nonumber\\
&&m_{\Sigma}(1,3,0)=m_A+\sqrt{2}h_p\frac{\bf
    \Phi_1}{3}+\sqrt{\frac{3}{5}}h_e<E>.
\label{compmass} 
\end{eqnarray}
Fixing the mass $m_A$, these formulas have the options of finetuning
two Yukawa couplings and four VEVs. If we get rid of ${210}_H$ we find
that both the triplet  mass $m_{\Sigma}(1,3,0)$ and the singlet
mass $m(1,1,1)$ can be made light by a single fine tuning. On the
other-hand if we use only  ${210}_H$, only  $m_{\Sigma}(1,3,0)$ can
be made light by a single fine-tuning. By the use of both ${54}_H$ and
${210}_H$ several options are available with a rich structure of
lighter fermion masses. 
In order to get both the triplet and the octet fermion masses light,
two finetunings are needed. 
A missing partner mechanism with two sets of fermion representations
${45}_F^{1,2}$ and a Higgs representation ${45}_H^Y$ has been used to
make the triplet fermionic DM light \cite{Frig-Ham:2010}.

%%%%%%%%%%%%%%%%%%%%%%%%%%%%%%%%%%%%%%%%%%%%%%%%%%%%%%%%%%%%%%%%%%%%%%%%%%%%%%%
\subsection{Triplet Fermion Dark Matter Phenomenology}
The phenomenology of a hyperchargeless triplet fermionic DM in the
non-SUSY model is similar
 to that of the wino DM in MSSM and SUSY
GUTs. This has been
extensively investigated recently \cite{cirelli:2005uq}
and also continues to be a subject of current importance
\cite{Hryzcuk:2014}. It is worthwhile to mention here different
constraints on their masses derived from direct and indirect searches
because of their relevance to the present model building. The even 
matter parity of fermion triplet DM $\Sigma_F(1,3,0)$, compared to odd (even)
matter parity of standard fermion (Higgs scalar), guarantees stability
of the DM by ruling out Yukawa interactions with SM particles. This
may make it difficult for the detection of the triplet fermionic DM
at the LHC and other hadron colliders. \\
\par\noindent{\bf (i).Triplet Fermion Mass from Relic Density:}\\
The only interaction of the
DM fermion with standard model particles is through gauge interaction
that leads to the well known mass difference
$m_{\Sigma^+}-m_{\Sigma^0}=166 $ MeV \cite{cirelli:2005uq}.
where we have denoted the mass of the charged (neutral) component of $\Sigma_F(1,3,0)$ as
$m_{\Sigma^+}(m_{\Sigma^0})$. Each of its two charged components has been estimated to be
heavier by $\sim 166$ MeV \cite {cirelli:2005uq}. Within  the
$3\sigma$ uncertainty, the observed DM relic abundance is $0.095 <
\Omega_{DM}h^2<0.125$ where $h=$ Hubble parameter.

For the triplet mass $m_{\Sigma}$ much larger than the $W$-boson mass,
the  Sommerfeld resonance enhancement plays a crucial role in the
annihilations of components of the $\Sigma_F$ leading to the observed
DM relic abundance.  Neglecting mass difference between the
charged and neutral components, the relevant cross
section taking into account the annihilation and co-annihilation of
all triplet components has been derived \cite{Frig-Ham:2010},
\begin{equation}
  <\sigma v> = \frac{37 g_{2L}^4}{96\pi m_{\Sigma}}, \label{anncross}
\end{equation}   
where $v=$ relative velocity of DM particles. The Sommerfeld enhancement enters
into the annihilation process because of the fact that the triplet
components are non-relativistic at the freezeout
temperature. Matching the theoretical prediction within the $3\sigma$
uncertainty of the observed value of the relic density $\Omega_{\rm DM}$ \cite{exptomegaDM}
results in the triplet mass $m_{\Sigma}= 2.75\pm 0.15$
TeV \cite{cirelli:2005uq, Hisano:2006nn,cirelli:2007xd} whereas a value
of $m_{\Sigma}=3.0 - 3.2$\,TeV has been also estimated
\cite{Hryzcuk:2014}.  A non-thermal production
of $\Sigma^0$ relic density due to the decay of color octet fermion,
$C_8 (8,1,0)_F$, has
been recently discussed in \cite{Aizawa:2014iea}.
Quite recently only the neutral components of DM candidates at the TeV scale
originating from RH fermionic triplets, rather than the LH triplets, have been suggested to be produced
at high temperature through non-equilibrium thermal
production process in non-SUSY SO(10) where the charged components
acquire larger intermediate scale masses
\cite{Mambrini:2015}.
The direct detection, indirect detection,
and collider search for triplet fermion DM at $p-p$ collider have been analysed 
in \cite{Cirelli:2014}. 
Phenomenology of  wino DM  in the mass range $500 -2000$ GeV
which has much similarity with this non-SUSY triplet fermionic DM, $\rho_3$, has been
also discussed recently \cite{Smohanty:2012}.
-

\par\noindent{\bf (ii). Direct Detection and Collider Signatures}\\
In general, for elastic scattering of a DM  particle (which is
electrically neutral) off nucleons either a
standard Higgs or a $Z$-boson exchange is needed in the t-channel of the dominant tree
diagrams. In the absence of such couplings of $\Sigma^0$, a sub-dominant
process occurs by the exchange of two virtual $W^{\pm}$ bosons in a
box diagram \cite{cirelli:2005uq}. This process leads to suppression
of spin independent cross section by $2-3$ orders below the
experimentally detectable value. However, such predicted cross sections
are measurable with improvement of detector sensitivities
\cite{Xenon:2009}. The inelastic scattering with a charged component
($\Sigma^+$ or $\Sigma^-$) is prevented because of kinematic
constraints since the mass difference, $m_{\Sigma^+}-m_{\Sigma^0}=166
$ MeV, is about three orders of magnitude above the kinetic energy of $\Sigma$
and also much above the proton-neutron mass difference, $m_n-m_p\sim
2$ MeV. If the triplet 
fermion  has mass $\sim 400$ GeV, its
contribution to the spin independent cross section is found to suffer
more deviation from the LUX direct bound \cite{Hisano:2014}.  
 
Prospects of observing signatures of the triplet fermion DM  at colliders have
been investigated in \cite{Hambye:2008,Aguila:2009,Roeck:2009,Cirelli:2014}. For $m_{\Sigma}\sim
2.7$ TeV and integrated luminosity of $100$fb$^{-1}$, the DM pair
production cross section at LHC in the channel $pp\to \Sigma\Sigma X$
has been shown to result in only one event \cite{Hambye:2008,Aguila:2009}. For better
detection capabilities upgradation of LHC with twice energy and more
luminosity has been suggested \cite{Roeck:2009} .

For detection at $e^+e^-$
collider that requires a collision energy of at least twice the DM mass, observation of
$\Sigma^+\Sigma^-$ pair production is predicted via $Z$ boson
exchange\cite{cirelli:2005uq,Hambye:2008}. The neutral pair $\Sigma^0\Sigma^{0*}$ can be
also produced, although at a suppressed rate, through  one-loop box
diagram mediated by two virtual $W$ bosons.  
After production such charged  components  would provide a clean signal as they would
manifest in long lived charged tracks due to their decays via
standard gauge boson interactions, $\Sigma^{\pm}\to W^{\pm} \to
\Sigma^0\pi^{\pm}$, or $\Sigma^{\pm}\to W^{\pm} \to
\Sigma^0 l^{\pm}\nu_l (l=e,\mu)$. The production of $e^{\pm}$ and
$\mu^{\pm}$ charged leptons but the absence of $\tau^{\pm}$ due to
kinematical constraint may be another distinguishing experimental
signature of the triplet fermionic DM. The decay length of such
displaced vertices is clearly predicted \cite{cirelli:2005uq,Hambye:2008} to be
$L_{\Sigma^{\pm}}\simeq 5.5$ cm. \\

A contrasting feature regarding the fate of the produced neutral component of the
triplet fermion DM,  $\Sigma^0 \subset$ SO(10) , different from the 
prediction of \cite{Hambye:2008,Aguila:2009}, has been observed in ref.\cite{Frig-Ham:2010}.
In the case of ref.\cite{Hambye:2008,Aguila:2009} it has been suggested that the
corresponding $\Sigma^0$ can decay into leptons.
But it has been noted in the context of the matter parity  conserved
SO(10) model \cite{Frig-Ham:2010} that
the decay product $\Sigma^0$  is stable because of its matter
parity. As such the production of this neutral component of the
triplet fermion DM
originating from SO(10) will be signalled through missing energy
\cite{Frig-Ham:2010}. This stability feature of  $\Sigma^0$ with its
TeV scale mass has
negligible impact on electroweak precision variables. 
These interesting features are applicable also in the
present model under investigation.

\subsubsection{Prospects from Indirect Searches}
PAMELA \cite{pamela} and FERMI/LAT \cite{fermi} experiments concluded the positron excess in case of the
WIMP as DM candidate which is again confirmed by recent AMS-02 \cite{ams} data \cite{Chen}. The electron and positron flux is still significant 
in the measurement of FERMI/LAT.
There are various constraints on the wino dark matter from different search channels such as 
antiprotons, leptons, dark matter halo from diffuse galactic gamma rays, 
 high latitude gamma-ray spectra, galaxy clusters, dwarf spheroids, 
 gamma-ray line feature, neutrinos from the galactic halo, CMB constraints, and
antideuterons \cite{Hryzcuk:2014}. 
In the case of the antiproton search channel the wino dark matter having mass close to the resonance i.e,
$2.4$ TeV, and thin zone of diffusion is  consistent with the antiproton measurement . 
The wino dark matter having mass near the resonance produces very small amount of leptons and large amount of positrons at
very low energy scale. This DM can not solve cosmic ray (CR) lepton puzzle because the lepton data can
rule out the very proximity of resonance. The galactic $\gamma$ rays impose a stringent limit on the 
wino DM model. With the inclusion of the $\gamma$ ray constraint, the limit on the wino DM changes. If the mass of  DM
 is $2.5$ TeV and it is in a thin diffusion zone, then it is excluded by the $\gamma$ ray data for a wide variation
of galactic CR propagation. There is also a very significant limit on the wino dark matter from high latitude
$\gamma$ ray spectra. For a $2.5$ TeV wino DM
the expected 10 year cross section is $1.5\times 10^{-25}$ $cm^3s^{-1}$ including DM substructures \cite{Hryzcuk:2014}. 
Possible signatures of DM annihilations are given from $\gamma$ ray observations\cite{Han,Hektor} towards nearby galaxy clusters 
 but observations in ref.\cite{HESS, MAGIC,Ajello,Dugger,Zimmer,Huang:12} have not seen any significant limits from $\gamma$ ray excess. The wino dark matter having mass
$2.4$ TeV can be ruled out in this search channel whereas all the other masses are allowed in the dwarf spheroids channel \cite{Hryzcuk:2014}. 
 The winos with masses heavier than 2 TeV are excluded by the HESS\cite{HESS} data
at $95\%$ CL. A new method to search for the indirect signals of DM annihilation is obtained due to
the motion of high energy neutrons towards the galactic center. Wino models
having the mass $2.4$ TeV can be observed in this search channel \cite{Hryzcuk:2014}. 
There is also a constraint on the wino dark matter due to the CMB temperature and polarization power
spectra. Taking WMAP-5 \cite{Komatsu} data and with $98\%$ CL,
the DM masses  in the region $2.3$ TeV to $2.4$ TeV have been excluded.
With WMAP-9 \cite{Hinshaw} the excluded limit is $2.25-2.46$ TeV. But
the combined search of WMAP-9 with ACT \cite{Sievers,Story} excludes the mass range of $2.18-2.5$
TeV . To search for the dark matter, the most
effective  channel is through antideuterons. 
Due to the smaller signal to back ground ratio at mass $2.5$ TeV, the resultant signal is very low with high uncertainty.
With the theoretical and experimental progress, there may be stringent limit on the wino dark matter \cite{Hryzcuk:2014}.\\

In our model the triplet fermionic thermal
DM resulting from any one of the nonstandard fermionic representations
${45}_F$, ${54}_F$, or ${210}_F$ would be adequate although we have
preferred to choose the minimal of these three representations  in order
to minimise the impact on GUT threshold uncertainties as discussed in
Sec.\ref{sec:th}.    

\section{Gauge Coupling Unification}\label{sec:uni}
In this section we discuss gauge coupling unification at the two-loop
level using lighter scalar and fermionic degrees of freedom motivated
by solutions to the neutrino masses by hybrid seesaw, dark matter and
leptogenesis.
 At first exact
unification of the three gauge couplings is
realized  using a triplet scalar $\Delta_L(1,3, 0)$ at
$M_{\Delta}=10^{12}$ GeV, a triplet
fermion $\Sigma_F(1,3,-1)$ at $M_T \sim 500-1000$ GeV, and, in
addition, a color octet fermion of Majorana-Weyl type at $M_{C_8}\sim
5 \times 10^{7}$ GeV. We
then estimate threshold effects on the GUT scale due to various superheavy
components in the theory. We discuss proton life prediction in the
model including these threshold uncertainties.

\subsection{ Unification with Lighter Fermions and Scalars}
 We use the standard renormalization group equations (RGEs) for the evolution
 of the three gauge 
 couplings \cite{GQW:1974} and their integral forms are 
\begin{eqnarray}
&&\frac{1}{\alpha_i(M_Z)}=\frac{1}{\alpha_i(M_{U})}
+\frac{a_i}{2\pi}{\rm ln}\left(\frac{M_{\Sigma}}{M_Z}\right)
+\frac{a'_i}{2\pi}{\rm ln}\left(\frac{M_{C_8}}{M_{\Sigma}}\right)
+\frac{a''_i}{2\pi}{\rm ln}\left(\frac{M_{\Delta}}{M_{C_8}}\right) \nonumber\\ 
&&+\frac{a'''_i}{2\pi}{\rm ln}\left(\frac{M_{U}}{M_{\Delta}}\right)  
+\Theta'_i+\Theta''_i+\Theta'''_i - \frac {\lambda_i}{12\pi},\label{tlrge}
\end{eqnarray}
where $M_{\Sigma}=$ triplet fermionic DM mass scale, $M_{\Delta}=$ LH
triplet mass mediating type-II seesaw, and $M_{C_8}=$ additional
fermion octet mass scale found to be necessary to achieve exact
  unification of the three gauge couplings at two-loop level. The one-loop
 coefficients $a_i^{'.''.'''}$ in their respective ranges of mass
 scales are shown in Table \ref{tab:beta} in the Appendix. The terms $\Theta'_i\,\,
 ,\Theta''_i$, and $\Theta'''_i$ are the two-loop contributions in the three
 different ranges of the mass scales with the respective coefficients 
$B_{ij}^{','','''}$ given in Table \ref{tab:beta}.

\begin{eqnarray}
&&\Theta_i=\frac{1}{4\pi}\sum_j B_{ij}{\rm ln}\frac{\alpha_j(M_{\Sigma})}{\alpha_j(M_Z)},
~~\Theta'_i=\frac{1}{4\pi}\sum_j B'_{ij}{\rm ln}\frac{\alpha_j(M_{C_8})}{\alpha_j(M_{\Sigma})},\nonumber\\
&&\Theta''_i=\frac{1}{4\pi}\sum_j B''_{ij}{\rm ln}\frac{\alpha_j(M_{\Delta})}{\alpha_j(M_{C_8})},
~~\Theta'''_i=\frac{1}{4\pi}\sum_j B'''_{ij}{\rm ln}\frac{\alpha_j(M_{M_U})}{\alpha_j(M_{\Delta})}.\label{thdef}
\end{eqnarray}

The term ${{\lambda_i}\over {12\pi}}$ represents GUT threshold effects
on the respective gauge coupling due to super-heavy particles existing
around $\mu=M_U$. These may be superheavy Higgs scalars, fermions, or
gauge bosons
\cite{Weinberg:1980,Hall:1981,Ovrut:1981,mkp-pkp:1991,rnm-mkp:1993,Langacker:1994}.

In terms of the experimentally determined parameters at the electroweak
scale \cite{PDG:2014}:$\sin^2\theta_W(M_Z)=0.23126 \pm 0.00005, \alpha
(M_Z)=1./127.9$ , and $\alpha_S(M_Z)=0.1187 \pm 0.0017$, we define   
\begin{eqnarray}
 &&P_S = \frac{2\pi}{\alpha(M_Z)}\left
 (1-\frac{8}{3}\frac{\alpha(M_Z)}{\alpha_S(M_Z)}\right), \nonumber\\
 &&P_{\Theta}= \frac{2\pi}{\alpha(M_Z)}\left
 (1-\frac{8}{3}\sin^2\theta_W(M_Z)\right),\label{pspth}
\end{eqnarray}
From the RGEs of eq.(\ref{tlrge}), the
corresponding RGEs for $P_S$ and $P_{\Theta}$ are obtained. These two are then solved to
yield formulas for the two mass scales $M_U$ and $M_{\Delta}$
\begin{eqnarray}
&&\ln\left(\frac{M_U}{M_Z}\right)=\frac{P_SB_{\Theta}-P_{\Theta}B_S}{D}+
\frac{C_{\Theta}B_S-C_SB_{\Theta}}{D}+\frac{B_ST_{\Theta}-B_{\Theta}T_S}{D},\nonumber\\
&&\ln\left(\frac{M_\Delta}{M_Z}\right)=\frac{A_SP_{\Theta}-A_{\Theta}P_S}{D}+
\frac{C_{S}A_{\Theta}-C_{\Theta}A_{S}}{D}+\frac{A_{\Theta}T_{S}-A_{S}T_{\Theta}}{D}.
\label{massrge}
\end{eqnarray}
In  eq.(\ref{massrge})
\begin{eqnarray}
&& A_S=(5/3)a_1^{'''}+a_{2}^{'''}-(8/3)a_3^{'''}, \nonumber\\
&& A_{\Theta}=(5/3)\left(a_1^{'''}-a_2^{'''}\right), \nonumber\\
&& B_S=(5/3)a_1^{''}+a_{2}^{''}-(8/3)a_3^{''}-A_S,\nonumber\\
&& B_{\Theta}=(5/3)\left(a_1^{''}-a_2^{''}\right)-A_{\Theta}, \nonumber\\
&& T_S=\frac{1}{6}\left[(8/3)\lambda_3-\lambda_2-(5/3)\lambda_1\right],
\nonumber\\
&& T_{\Theta}=\frac{5}{18}\left[\lambda_2-\lambda_1\right],\nonumber\\
&& D=A_SB_{\Theta}-A_{\Theta}B_S.\label{paraeq}
\end{eqnarray}
Apart from depending upon the RG coefficients, the quantities $C_S$
ans $C_{\Theta}$ in eq.(\ref{massrge}) depend upon the lighter mass
scales $M_{\Sigma}$ and $M_{C_8}$
\begin{eqnarray}
&& C_S=\left[(5/3)(a_1^{'}-a_1^{''}) +a_{2}^{'}-a_{2}^{''}-(8/3)(a_3^{'}-a_3^{''})\right]{\rm
      ln}({M_{C_8}\over M_Z}),\nonumber\\ 
&&+\left[(5/3)(a_1-a_1^{'}) +a_{2}-a_{2}^{'}-(8/3)(a_3-a_3^{'})\right]{\rm
      ln}({M_{\Sigma}\over M_Z}),\nonumber\\
&& C_{\Theta}=\left[(5/3)(a_1^{'}-a_2^{'}
    -a_{1}^{''}+a_{2}^{''})\right]{\rm ln}({M_{C_8}\over M_Z}),\nonumber\\ 
&& +\left[(5/3)(a_1 -a_2 -a_{1}^{'}+a_{2}^{'})\right]{\rm ln}({M_{\Sigma}\over M_Z}).
\label{cscth}
\end{eqnarray}
In deriving the analytic formulas in eq.(\ref{massrge}) we have
ignored the two-loop terms for the sake of simplicity although they
have been included in numerical estimations of mass scales involved. 
It is clear that in  eq.(\ref{massrge}) the first two terms in the
RHS for the two mass scales $M_U$ and $M_{\Delta}$ represent the
one-loop contributions but the third term in each case represents the
corresponding threshold correction.

\begin{figure}[h!]
\begin{center}
\includegraphics[scale=1.5]{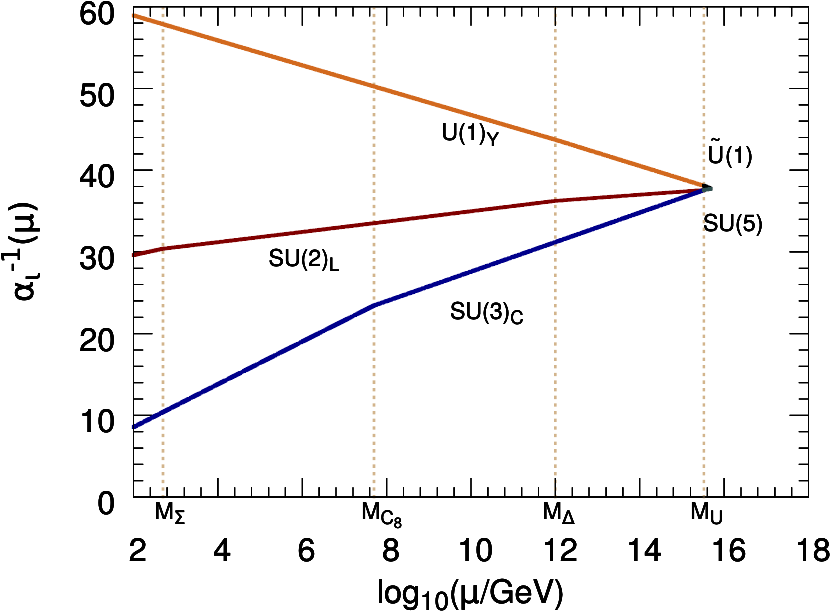}
\end{center}
\caption{ Unification of couplings of the SM gauge group
  $SU(3)_C\times SU(2)_L\times U(1)_Y$ in the presence of LH triplet
  scalar $\Delta_L$, the triplet fermionic dark matter $\Sigma_F$, and the color
  octet fermion $C_8$ as described in the text. The ordinates
  corresponding to these masses $M_{\Sigma},M_{C_8}, M_{\Delta}$ and
  the GUT scale $M_U$ are indicated along the X-axis.}
  the 
\label{fig:cu}
\end{figure}

At
first retaining only one-loop and the two-loop contributions we
 find excellent unification of the three gauge couplings for
 $M_{\Sigma}=500-1000$ GeV,  $M_{C_8} \sim 5\times 10^{7}$ GeV and $M_{\Delta}=
10^{12}$ GeV. This is shown in Fig.\ref{fig:cu}. 

In this model we have found the necessity of either two color octet scalars
$S_8(8,1,0)$  or a single octet fermion $C_8(8,1, 0)$ at mass
$M_{C_8}\sim 5\times 10^{7}$ GeV,
in addition to the triplet fermionic  DM candidate $\Sigma_F(1,3,0)$ and
the LH triplet scalar $\Delta_L(1,3,-1)$. This color octet fermion is thus
safely above the cosmologically allowed limit \cite{arkani}.
 The two-loop prediction of the GUT scale and the gauge coupling are
 \begin{eqnarray}
 &&M_U^0= 10^{15.56} {\rm GeV} ,\nonumber\\
&&g_G(M_U)=0.573\label{MU} 
\end{eqnarray}

\section{Threshold Corrections and Proton Lifetime Prediction}\label{sec:th}

\subsection{Threshold Effects on the GUT Scale}
As pointed out in Sec.\ref{sec.1}, the superheavy components of the 
representation ${126}_H$ is expected to contribute substantially to the GUT
threshold effects on the GUT scale $M_{U}$ and hence on the proton
lifetime predictions. In this
estimation at first we assume all the superheavy DM components in ${45}_F$ to
be exactly degenerate with the GUT scale leading to their vanishing
threshold effects. In the next step we estimate the fermionic
contribution by following the same procedure \cite{Weinberg:1980,Hall:1981,Ovrut:1981,Langacker:1994,mkp-pkp:1991}.

From the last term in eq.(\ref{massrge}), the analytic formula for
GUT threshold effects on the unification scale is 
\begin{eqnarray} 
&&\Delta {\rm ln}(M_U/M_Z)=(54/1829)\left[(40/81)\lambda_1-
    (4/27)\lambda_2-(28/81)\lambda_3\right]\label{mumzth}
 \end{eqnarray}
where for the ith super-heavy scalar component
$\lambda_i=tr(t_i^2)ln(M_{S_i}/M_U)$. But for Weyl (Dirac) fermions 
near the GUT scale there is  multiplicative factor $4(8)$. The
numerical values for $tr(t_i^2)$ for each submultiplet has been given
in the corresponding Tables in Appendix B.
 
We next evaluate the functions $\lambda_i(M_U)$ involving small logs
caused due to super-heavy scalar components in the loop. These are
contained  in the SO(10) Higgs
representations ${10}_H,{45}_H$, and ${126}_H$. We
further introduce the ``partially degenerate'' assumption on the
super-heavy component masses of Higgs scalars which has been found to
be useful in handling large representations especially in $SO(10)$
\cite{rnm-mkp:1993}. Under this assumption all super-heavy scalar
masses belonging to a given representation have a common degenerate
mass. Then using
decompositions of representations shown in the Appendix we find
\begin{eqnarray}
&&\lambda_1=17/5+4\eta_{(10)}+(0)\eta_{(45)}+136\eta_{(126)},\nonumber\\
&&\lambda_2=6+4\eta_{(10)}+2\eta_{(45)}+140\eta_{(126)},\nonumber\\
&&\lambda_3=8+4\eta_{(10)}+3\eta_{(45)}+140\eta_{126},\label{lambthi}
\end{eqnarray}
where $\eta_X={\rm ln}(M_X/M_U)$. The constant terms on the RHS of
eq.(\ref{lambthi}) represent the contributions of $33$ super-heavy gauge
bosons assumed to be degenerate at the GUT scale $M_U$. The dominant
contributions to the threshold factors $\lambda_i$ in
eq.(\ref{lambthi}) arising out of the super-heavy scalar
components of ${126}_H$ are quite explicit.

 Using eq.(\ref{lambthi})
in eq.(\ref{mumzth}) and maximizing the uncertainty \cite{rnm-mkp:1993} gives 
\begin{eqnarray}
&& [\frac{M_U}{M_U^0}]_{S}=10^{\pm 0.928\eta_S}, \nonumber\\
&& \eta_S=|{\rm log}_{10}\left[\frac{M_{SH}}{M_{U}}\right]|,\label{muthS}
\end{eqnarray}
where $M_{SH}$ is the super-heavy Higgs mass scale and $M_U^0$
represents the two-loop solution of eq.(\ref{MU}) without threshold
effects. 
Similarly excluding the light triplet DM component $\Sigma_F(1,3,0)$, the rest of the fermionic component of the representation
${45}_F$ contribute to the threshold effects
\begin{eqnarray}
&& [\frac{M_U}{M_U^0}]_{F}=10^{\pm 0.253\eta_F}, \nonumber\\
&& \eta_F=|{\rm log}_{10}\left[\frac{M_F}{M_{U}}\right]|,\label{muthF}
\end{eqnarray} 
We also note that the degenerate super-heavy gauge bosons contribute a
very small correction with a positive sign
\begin{equation}
 [\frac{M_U}{M_U^0}]_{V}=10^{ 0.0227}. \label{muthV}
\end{equation} 
In general
following Coleman-Weinberg \cite{Coleman-Weinberg} idea, $M_{SH}$ could vary quite naturally
within the range $M_U/10$ to $10 M_U$. As the the super-heavy fermionic
components are unaffected by such corrections it may be natural to
treat their masses to be degenerate at the GUT scale or at a
degenerate mass $M_F$ around $M_U$. In the first case they do not
contribute to 
threshold corrections to the 
corrected unification scale. We have considered the general case with
degenerate mass $M_F=(1/10 \to 10)M_F$. Adding all corrections
together we get
\begin{equation}
M_U=10^{15.56 +0.0227\pm 0.928\eta_S\pm 0.253\eta_F} {\rm GeV} \label{MUtotal}
\end{equation}   
Treating this as the mass of super-heavy gauge bosons mediating proton decay, we
next estimate proton lifetime prediction in the model.
\subsection{Proton Lifetime Prediction}
As the unification scale predicted by this model has an uncertainty
naturally dictated by the matter parity motivated SO(10) model, it
would be interesting to examine its impact on proton life time predictions 
for  $p \to e^+\pi^0$
for which there are ongoing dedicated experimental searches \cite{Babu-Pati:2010,Nath:2007,nishino:2009,nishino:2012}
with measured value of the lower limit on the life time
\cite{Shiozawa:2013,Hyper K.}\\
\begin{eqnarray}
&&\tau_p^{expt.}~\ge ~1.4\times 10^{34}~~{\rm yrs.} \label{taupexpt}        
\end{eqnarray}
 Including strong and electroweak renormalization
effects on the ${\rm d}=6$ operator and taking into account quark mixing, chiral symmetry breaking
effects, and lattice gauge theory estimations, the decay rates
 for the two models are \cite{Babu-Pati:2010,Buras:1978,bajc}, 
\begin{eqnarray}   
&&\Gamma(p\rightarrow e^+\pi^0) \nonumber\\ 
&&=\frac{m_p}{64\pi f_{\pi}^2}
\frac{{g_G}^4}{{M_U}^4})|A_L|^2|\bar{\alpha_H}|^2(1+D'+F)^2\times R,\nonumber\\
\label{width}
\end{eqnarray}
where $ R=[A_{SR}^2+A_{SL}^2 (1+ |{V_{ud}}|^2)^2]$ for $SU(5)$, but $R=
[(A_{SR}^2+A_{SL}^2) (1+ |{V_{ud}}|^2)^2]$ for $SO(10)$, $V_{ud}=0.974=$ 
 the  $(1,1)$ element of $V_{CKM}$ for quark mixings, and
$A_{SL}(A_{SR})$ is the short-distance renormalization factor in the
left (right) sectors.  In eq.(\ref{width}) $A_L=1.25=$
long distance renormalization factor but  
$A_{SL}\simeq A_{SR}=2.542$. These are numerically estimated by
evolving the ${\rm dim.} 6$ operator for proton decay by using the
anomalous dimensions of ref.\cite{Buras:1978} and the beta function
coefficients for gauge couplings of this model. In eq.(\ref{width})  
 $M_U=$ degenerate mass of  super-heavy gauge bosons, $\bar\alpha_H =$
hadronic matrix elements, $m_p =$ proton mass
$=938.3$ MeV, $f_{\pi}=$ pion decay 
constant $=139$ MeV, and the chiral Lagrangian parameters are $D=0.81$ and
$F=0.47$. With $\alpha_H= \bar{\alpha_H}(1+D'+F)=0.012$ GeV$^3$ estimated from 
lattice
gauge theory computations  \cite{Aoki:2007}, we obtain  $A_R \simeq A_LA_{SL}\simeq
A_LA_{SR}\simeq 3.18$ and the expression for the
 inverse
decay rate is,   

\begin{eqnarray}
&&\Gamma^{-1}(p\rightarrow e^+\pi^0) \nonumber\\
&& =
  \frac{4}{\pi}\frac{f_{\pi}^2}{m_p}\frac{M_U^4}{\alpha_G^2}\frac{1}{\alpha_H^2
    A_R^2}\frac{1}{F_q},\label{taup}
\end{eqnarray}
where the GUT-fine structure constant $\alpha_G=0.0263$ and the
factor  $F_q=2(1+|V_{ud}|^2)^2\simeq 7.6$ for $SO(10)$. 
This formula  reduces to
the form given in \cite{psb:2010,Babu-Pati:2010} and sets the lower limit for the non-SUSY
GUT scale to be $M_U\ge 10^{15.5}$ GeV from the lower limit
 of eq.(\ref{taupexpt}).

\begin{figure}[h!]
\begin{center}
\includegraphics[scale=0.6]{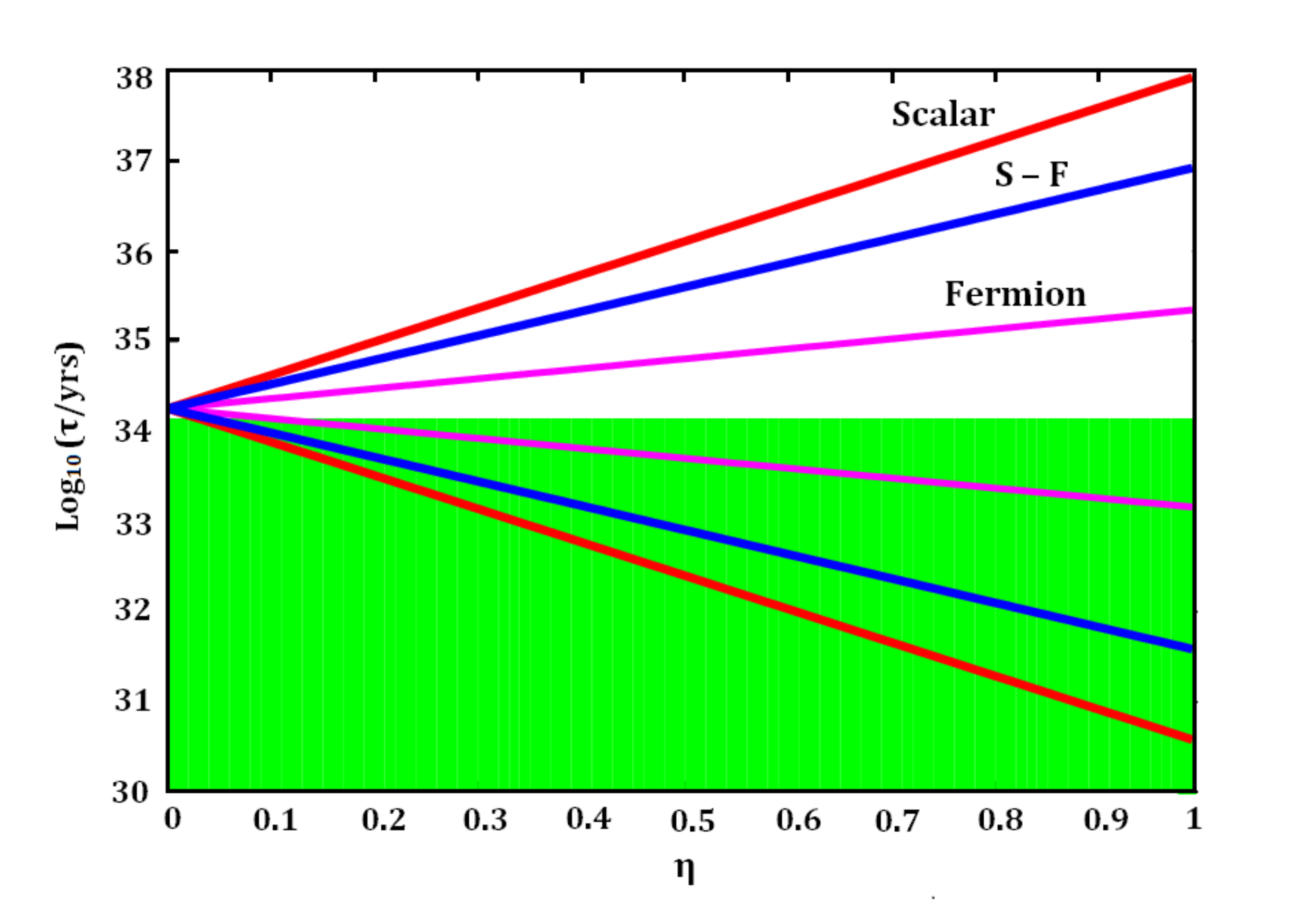}
\end{center}
\caption{Proton lifetime prediction for the decay mode $p\to e^+\pi^0$
shown by slanting solid lines as a function of
${\eta} =\eta_S(\eta_F)
=|{\log}_{10}(M_{SH}/M_U)|(|{\log}_{10}(M_F/M_U)|)$ for super-heavy
scalar(fermion) components. The shaded green colored region is
ruled out by the current experimental bound. The point at $\eta_S=\eta_F =0$ represents the
model prediction at two-loop level without threshold effects with
$\tau_P^0=1.8\times 10^{34}$ yrs.} 
\label{fig:taup}
\end{figure}

 Now using  the estimated values of the model parameters  eq.(\ref{taup}) gives,
\begin{eqnarray}
&&\tau_p^{SO(10)}\simeq 1.8\times 10^{34\pm 3.712\eta_S\pm
    1.012\eta_F}~~{\rm yrs}.\label{taupnum}
\end{eqnarray}
 As an example, a super-heavy scalar mass splitting by
a factor $2(1/2)$ from the GUT scale gives  $\eta_S=0.3(-0.3)$ leading to $\tau_p \sim 1.8\times
10^{34\pm 1.11}$~~yrs even if all fermion masses are at
$M^0_U$. Similarly if all super-heavy scalar masses are degenerate at
the unification scale $M^0_U$, the super-heavy fermions with their mass
splitting factor $2(1/2)$ lead to   $\tau_p \sim 1.8\times
10^{34\pm 0.3}$~yrs. These
lifetimes  are clearly above the current experimental limit but
accessible to ongoing searches. The proton lifetime predictions as a
function of $\eta=\eta_S$ or $\eta=\eta_F$  are
shown in Fig.\ref{fig:taup} for the $p\to e^+\pi^0$ decay mode.

It is clear that most of the uncertainties arise out of the GUT threshold
corrections due to  the larger Higgs representation ${126}_H$  which
plays the central role in determining the contents of dark matter and their
stability in the non-SUSY SO(10) by preserving matter parity as gauged
discrete symmetry. We note that such uncertainties which are crucial
for proton decay searches have been
estimated here for the first time. The DM motivated SO(10) also
predicts additional threshold corrections to proton lifetime
predictions especially due to fermions. Although this may enhance the uncertainty further, in one
class of solutions the
model also offers an interesting new possibility compared to GUTs without
fermionic dark matter. The fermionic threshold corrections may contribute to
cancel out a substantial part of the scalar threshold effects  in another class of solutions
which are shown by the blue curve marked $S-F$ in Fig.\ref{fig:taup}. With
this cancellation, the proton decay has somewhat more probability for
detection by the ongoing  searches.

\section{Summary and Conclusion}\label{sec:conc}
In this work we have attempted unification of gauge couplings of the
non-SUSY standard gauge theory by addressing solutions to three
of its outstanding problems: neutrino masses, dark matter, and baryon
asymmetry of the Universe (BAU). To achieve
these objectives we have exploited an interesting breaking pattern of 
non-SUSY SO(10) by assigning GUT scale VEV to the representation ${126}_H$ where matter parity is conserved as a natural gauged
discrete symmetry of the SM that guarantees dark matter stability. As
the origin  of dark matter candidates, the model classifies 
non-standard fermionic or scalar representations of non-SUSY SO(10)
 carrying even or odd matter parity containing suitable components of dark
matter. It predicts the type-I $\oplus$ type-II as the hybrid
seesaw formula for neutrino masses driven by LH scalar triplet
$\Delta_L (1, 3,-1)$ and heavy RH neutrinos. This 
formula  has been  used here to
fit the neutrino oscillation data that predicts the heavy masses of the
scalar triplet and the RH$\nu$ masses. We have carried out this
fitting procedure using values of the Dirac neutrino mass matrix
derived in two ways by assuming $u$-quark diagonal or the $d$-quark diagonal bases. For a given
intermediate mass value  of the scalar triplet, induced VEV, and Dirac
neutrino mass matrix, this seesaw
formula being quadratic in Majorana neutrino Yukawa coupling $f$,
predicts two distinct sets of RH$\nu$ masses:(i)Compact spectrum where all
three masses are heavier than the Davidson-Ibarra (DI) bound, and (ii)
Hierarchical spectrum where only $N_1$ is lighter than the DI bound.
These solutions provide a variety of results on 
 the surviving lepton asymmetries after washout factors are adequately
 taken into account. We have
carried out a complete flavor analysis of the RH$\nu$ decays and
exploited solutions to Boltzmann equations in every case to arrive at
the model 
predictions  on the baryon asymmetry. Although the decay of the LH scalar triplet in this model
is found to yield negligible CP-asymmetry, it contributes quite
significantly through the new Feynman diagram it generates for the
vertex correction of RH$\nu$ decays. In fact this contribution to
the CP-asymmetry is found to be as dominant as other contributions
without triplet mediation. 
The decay of the lightest RH$\nu$ in the compact spectrum scenario predicts the
values of BAU in agreement with the existing data when the Dirac
neutrino mass determination is associated with either the $u$-quark
diagonal basis
or the $d$-quark diagonal basis. In the case of hierarchical  spectrum of RH neutrinos, the
right value of BAU is predicted by the $N_2$ decay where  the Dirac
neutrino mass is associated with the $u$-quark diagonal basis. This has been found possible even if the initial condition satisfies vanishing $N_2$ abundance.\\ 

 With the matter parity available as the stabilising
discrete symmetry for dark matter, the neutral component of
hyperchargeless triplet fermion $ \Sigma_F(1,3,0) \subset
{45}_F\subset SO(10)$ having even matter parity is well accommodated as
a candidate for thermal
dark matter  at TeV scale whose phenomenology has been discussed extensively in the 
literature and summarized here. Having thus addressed solutions to the three outstanding
problems of the SM as stated above, we implemented unification of the three gauge couplings
which needed a fermionic color octet of mass $M_{C_8}\sim 5\times 10^{7}$
GeV, in addition to the heavy Higgs scalar triplet, and the fermionic
triplet dark matter. The two-loop solutions yielded excellent
unification with the predicted GUT scale value $M_U=10^{15.56+0.0288}$
GeV  where the  small positive fraction in the exponent is due to degenerate
masses of all superheavy gauge bosons at $M_U^0$ that causes nearly
$30\%$ increase in the proton lifetime prediction over its two-loop prediction.
 Noting the 
compelling requirement of the scalar representation ${126}_H$ to drive
the symmetry breaking in this SO(10)
model, its superheavy components predict
substantial GUT threshold effects on the unification scale and proton
lifetime. We have also estimated threshold corrections on the
predicted proton lifetime due to superheavy fermions in ${45}_F$. An
interesting possibility of cancelling out a substantial part of
threshold
 corrections due to
scalars by fermions has been pointed out. We find that a large region
of the parameter space can be explored by the ongoing searches on proton
decay $p\to e^+\pi^0$.
 
In conclusion we find that in the non-supersymmetric standard gauge
theory, the predictions for neutrino masses, dark matter, baryon asymmetry of the
universe, unification of gauge couplings, and proton lifetime
accessible to ongoing searches can be successfully implemented
through direct breaking of non-SUSY SO(10) with particle content
inherent to  matter parity
conservation. The only additional particle needed beyond these
requirements for coupling unification is a color octet Weyl fermion (or a pair of complex color octet scalars) which also belong
to the SO(10) GUT representation. The introduction of the scalar
triplet $\Delta_L$ at the intermediate scale brought in naturally by
matter parity conservation in SO(10) causes remarkable changes in the
model predictions over its conventional values. The very fact of successful
implementation of the current programme in SO(10) resolves the issue
of parity violation as a monopoly of weak interaction.    
   
\section{Appendix: Renormalization Group Coefficients for Unification
of Gauge Couplings and Threshold Uncertainties}\label{sec.app}
In the Appendix A below we provide various decompositions of SO(10)
representations under different subgroups relevant for the present
work. In Appendix B we give different beta function coefficients along
with particle content for different mass scales.
\subsection{Appendix A: Decomposition of Representations and Beta
  Function Coefficients}
In this Appendix we present decompositions of non-SUSY SO(10) representations
under  SU(5) as shown in Table \ref{tab:decomp}.

\begin{table}[h!]
\centering
\begin{tabular}{|l|}
\hline
SO(10) $\supset$ SU(5)\\ 
10 $\supset$ 5 +  ${5}^{\dagger}$ \\
16 $\supset$ 10 + ${5}^{\dagger}$+ 1 \\
45 $\supset$ 10 + ${10}^{\dagger}$ + 1 + 24 \\
54 $\supset$ 24 + 15 + ${15}^{\dagger}$ \\
120$\supset$ 5 +${5}^{\dagger}$+${10}^{\dagger}$+${45}^{\dagger}$
+10 + 45 \\
126 $\supset 5^{\dagger}$+ 45+${15}^{\dagger}$ 
+${50}^{\dagger}$ + 10 + 1\\
210 $\supset$ 1+ 24 +  ${10}^{\dagger}$+ 10 + 40 + ${40}^{\dagger}$\\
 \hspace{1cm} + 75 + 5 +${5}^{\dagger}$\\
\hline
\end{tabular}
\caption{Decomposition of SO(10) representations into  SU(5) representations
 \cite{Fukuyama:2004ps}.}
\label{tab:decomp}
\end{table}

\subsubsection{ Particle Content and Beta Function Coefficients}
In this subsection we present the particle  content used in various
ranges of mass scales as shown in Table \ref{tab:particles} and the corresponding beta-function coefficients
which have contributed for the gauge coupling unification,
leptogenesis, and dark matter as shown in Table \ref{tab:beta}. 

\begin{table}[h!]
\centering
\begin{tabular}{|l|l||l|}
\hline 
Energy Scale & Particle content  \\
\hline
$M_Z-M_T$ & SM Particles   \\ 
\hline
$M_T-M_{O}$ & SM$+(1,3,0)_F$ \\
\hline
$M_{O}-M_{\Delta}$ & ${\rm SM}+(1,3,0)_F +(8,1,0)_F$ \\
\hline
$M_{\Delta}- M_{U}$ & ${\rm SM}+(1,3,0)_F+(8,1,0)_F +(1,3,1)_H$ \\
\hline
\end{tabular}
\caption{Particle content of the model in different ranges of mass scales.}\label{tab:particles}
\end{table}

\begin{table}[h!]
\centering
\begin{tabular}{|c|c|c|c|c|}
\hline
&\multicolumn{2}{|c|}{Model} \\
\hline
$\mu$ & $a_i$ & $a_{ij}$  \\
\hline
$M_Z-M_T$&$\bpmat 41/10 \\ -19/6 \\ -7\epmat$ & 
$\bpmat
199/50 &27/10&44/5\\
9/10 &35/6 & 12\\
11/10& 9/2 &-26\\
\epmat$
\\ 
\hline
$M_T-M_{O}$ &$\bpmat 41/10 \\ -11/6 \\ -7\epmat$ & 
$\bpmat
199/50 &27/10&44/5\\
9/10 &163/6 & 12\\
11/10& 9/2 &-26\\
\epmat$
\\ 
\hline
$M_{O}-M_\Delta$ & $\bpmat 41/10 \\ -11/6 \\ -5\epmat$ & 
$\bpmat
199/50 & 27/10 & 44/5\\
9/10 & 163/6 & 12\\
11/10& 9/2 & 22\\
\epmat$
 \\ 
\hline
$M_\Delta-M_U$ & $\bpmat 43/10 \\ -7/6 \\ -5\epmat$ & 
$\bpmat
83/10 & 171/10 & 44/5\\
57/10 & 275/6 & 12\\
11/10& 9/2 & 22\\
\epmat$
\\ 
\hline
\end{tabular}
\caption{One-loop and two-loop beta function coefficients in the respective ranges of
mass scales.}\label{tab:beta}
\end{table}
\newpage
\subsection{Appendix B:~~Super-heavy Particles and Coefficients for
  Threshold Effects}
In this subsection we identify the super-heavy particle contents of
various SO(10) representations with their quantum numbers and beta
function coefficients under the SM gauge group. These coefficients
shown in Table \ref{tab:10}, Table \ref{tab:45}, and Table \ref{tab:126b}
have been used for the estimation of threshold effects on proton
lifetime predictions. \\
   
\begin{table}[h!]
\begin{center}
\begin{tabular}{|c|c|c|}
\hline \hline
$SU(5)$ & $\left(3_C,2_L,1_{Y} \right)$ & tr($t_i^2$) \\
\hline
${\bf 5}$  & $\left({\bf 3,1};-\frac{1}{3} \right)$ & (1, 0, 2/5)\\
 & $\left({\bf 1,2};-\frac{1}{2} \right)$ & (0, 1, 3/5)\\
 \hline
${\bf\overline{5}}$& $\left({\bf \overline{3},1};\frac{1}{3} \right)$ & (1, 0, 2/5)\\
 & $\left({\bf 1,2};\frac{1}{2} \right)$ & (0, 1, 3/5)\\
\hline \hline
\end{tabular}
\end{center}
\caption{Decomposition of the complex ${\bf 10}$ representation under SU(5) and 
one-loop coefficients.}
\label{tab:10}
\end{table}

\begin{table}[h!]
\begin{center}
\begin{tabular}{|c|c|c|}
\hline \hline
$SU(5)$ & $\left(3_C,2_L,1_{Y} \right)$ & tr($t_i^2$) \\

\hline
$\left({\bf10} \right)$ & $\left({\bf 1,1};-1 \right)$ & (0, 0, 3/5) \\
&  $\left({\bf 3,2};-\frac{5}{6} \right)$ & (1, 3/2, 5/2) \\
& $\left({\bf \overline{3},1};-\frac{2}{3} \right)$ & (1/2, 0, 1/5) \\
\hline
$\left({\bf \overline{10}} \right)$ & $\left({\bf 1,1};1 \right)$ & (0,0,3/5)\\
& $\left({\bf \overline{3},2};\frac{5}{6} \right)$ & (1, 3/2, 5/2)\\
& $\left({\bf 3,1};\frac{2}{3} \right)$ & (1/2, 0, 1/5) \\
\hline
$\left({\bf 24}\right)$ & $\left({\bf 1,1};0 \right)$ & (0, 0, 0) \\ 
& $\left({\bf 1,3};0 \right)$ & (0, 2, 0) \\
& $\left({\bf 8,1};0 \right)$ & (3, 0, 0) \\  
& $\left({\bf 3,2};\frac{1}{6} \right)$ & (1, 3/2, 1/10) \\
& $\left({\bf \overline{3},2};-\frac{1}{6} \right)$  & (1, 3/2, 1/10) \\
\hline \hline
\end{tabular}
\end{center}
\caption{Decomposition of the real ${\bf 45}$ representation under
  SU(5) and one-loop coefficients. For the sake of convenience, the would-be goldstone modes of all
super-heavy gauge bosons have been provided from the scalar
representation ${45}_H$.}
\label{tab:45}
\end{table}
\newpage

\begin{table}[h!]
\begin{center}
\begin{tabular}{|c|c|c|}
\hline \hline
$SU(5)$  & $\left(3_C,2_L,1_{Y} \right)$ & tr($t_i^2$)\\
\hline
$\left({\bf 5}\right)$ & $\left({\bf 3,1};-\frac{1}{3} \right)$ & (1, 0, 2/5)\\
& $\left({\bf 1,2};-\frac{1}{2} \right)$ & (0, 1, 3/5) \\
\hline
$\left({\bf 15}\right)$ & $\left({\bf 6,1};\frac{2}{3} \right)$ & (5, 0, 16/5) \\
& $\left({\bf 3,2};\frac{1}{6} \right)$ & (2, 3, 1/5) \\
& $\left({\bf 1,3};1 \right)$ &  (0, 4, 18/5)\\
\hline
$\left({\bf \overline{10}}\right)$  & $\left({\bf 1,1};-1 \right)$ & (0, 0, 6/5)\\	 
& $\left({\bf 3,1};-\frac{2}{3} \right)$ & (1, 0, 8/5)\\
& $\left({\bf \overline{3},2};-\frac{1}{6} \right)$ & (2, 3, 1/5) \\
\hline
$\left({\bf 50}\right)$ & $\left({\bf \overline{6},3};-\frac{1}{3} \right)$ & (15, 24, 12/5)\\
& $\left({\bf 1,1};0 \right)$ & (0, 0, 0)\\
& $\left({\bf 3,1};-\frac{1}{3} \right)$ & (1, 0, 2/5) \\
& $\left({\bf 6,1};-\frac{2}{3} \right)$ & (5, 0, 16/5)\\
& $\left({\bf \overline{3},2};-\frac{1}{6} \right)$ & (2, 3, 1/5) \\
& $\left({\bf 8,2};-\frac{1}{2} \right)$ & (12, 8, 24/5)
\\
\hline
$\left({\bf \overline{45}} \right)$ & $\left({\bf \overline{3},1};\frac{1}{3} \right)$ & (1, 0, 2/5)\\
& $ \left({\bf \overline{3},3};\frac{1}{3} \right)$ & (3, 12, 6/5)\\ 
& $\left({\bf 3,1};\frac{2}{3} \right)$ & (1, 0, 8/5)\\
& $\left({\bf 1,2};\frac{1}{2} \right)$ & (0, 1, 3/5) \\
& $\left({\bf 6,1};\frac{1}{3} \right)$ & (5, 0, 4/5)\\
& $\left({\bf 3,2};\frac{1}{6} \right)$ & (2, 3, 1/5)\\ 
& $\left({\bf 8,2};\frac{1}{2} \right)$ & (12, 8, 24/5)\\
\hline \hline
\end{tabular}
\caption{Decomposition of the representation ${\bf \overline{126}}$
  under SU(5) and one-loop coefficients}
\label{tab:126b}
\end{center}
\end{table}

\subsection{ Appendix C: A discussion on charged fermion mass parametrization}
While all single step descents of SUSY GUTs leading to MSSM exhibit
almost profound gauge coupling unification,
there has been several attempts in  SUSY SO(10) to explain  fermion
masses of three generations of quarks and leptons along with the
attractive phenomena like $b-\tau$ or $t-b-\tau$ Yukawa
unification. In certain other cases  approximate validity of some of
the  Georgi-Jarlskog
\cite{GJ:1979} type  mass relations
\bea
m^0_{\mu}&\approx& 3m^0_{s},\nonumber\\
m^0_{\tau}&\approx& m^0_{b},\nonumber\\
m^0_{d}&\approx& 3m^0_{e}.\label{GJrel}
\eea
have been found to hold at the GUT scale.
  While some recent works have presented 
 very attractive  details of data analysis with  $\chi^2$-fit
 \cite{Joshipura:2011} as pointed out in Sec.1, a much larger number of other
 research papers have confined to partially quantitative  or  qualitative
 representations of the charged fermion masses as these latter types
 of investigations focus on other challenging issues of particle physics.
Compared to such interesing results on fermion mass fits in the direct
breaking model of SUSY SO(10)
\cite{Joshipura:2011}, non-SUSY models need at least one  intermediate
gauge symmetry to ensure gauge coupling unification within the
constraint of extended survival hypothesis \cite{del Aguilla:1981,rnm-gs:1983}.
 Also unlike the MSSM or SUSY SO(10), the RG extrapolated values of charged fermion
 masses through either SM or two-Higgs doublet model in the bottom-up
 approach \cite{dp:2001}  do not exhibit a precise $b-\tau$
 Yukawa unification at the scale $\mu \sim 10^{16}$ GeV. Unlike the attempts to present all fermion masses in SUSY SO(10)
through $\chi^2$ fit and  non-SUSY case with $SU(4)_C\times SU(2)_L
\times U(1)_R$ intermediate symmetry \cite{Joshipura:2011}, to our knowledge no such
analysis appears to have been done so far in the direct breaking of
non-SUSY SO(10) where gauge coupling unification itself under the minimal fine-tuning constraint
\cite{del Aguilla:1981,rnm-gs:1983} is highly challenging.
 In attempts to confront more challenging problems in SUSY or
non-SUSY SO(10), a
number of recent works have  ignored the
question of fitting the charged fermion masses 
while confining mainly to only neutrino masses and mixings, or at
most a qualitative presentation of charged fermion masses \cite{Five dim:2015,Fukuyama:2016,mkp:2008,nofit0,nofit1,nofit2,nofit3,nofit4,Malinsky-Romao-Valle:2005,PSB-rnm:2010,Babu-Pati:2010,Fukuyama:Axion:2005,Romao:2015,Hisano:Zp:2016,Bobby:2016,Fukuyama:Infl:2005,Ellis:2016,Shafi:2016}. However,
even though a $\chi^2$ fit \cite{Joshipura:2011} is not our present
goal, we point out how the charged fermion masses may be parameterized
 within this direct breaking model of non-SUSY
SO(10) while successfully encompassing standard model paradigm at
lower scales, neutrino masses, baryon
asymmetry, dark matter, gauge coupling unification , and GUT scale
parity restoration.\\

The Higgs representations  ${10}_H, {126}_H$ , and ${120}_H$
are known to contribute                                            
to fermion masses through the corresponding renormalizable Yukawa interactions.
We include two copies of ${10}_H$ fields in the corresponding renormalizable part of the Yukawa Lagrangian 
\begin{equation}
-{\cal L}^{(10)}=\sum_{p=u,d}Y^{(p)}_{ij} {16}_i {16}_j {10}_{H_p}, \label{Yuk10}
\end{equation}
 The Yukawa term
$f 16. 16. {126}_H$ has been found to be specifically suitable in
approximately satisfying the GJ type relations  in the
down quark and charged lepton sectors. Conventionally, the same matrix $f$ also contributes
to the RH neutrino mass matrix $M_N=f v_R$  which plays a crucial role
in the type-I and type-II seesaw components of the hybrid seesaw formula used in
this work. Therefore, the prime concern for charged fermion mass fit in the
present model may be the
 smallness of the value of the
 matrix elements  $f_{ij}\sim {\cal O}(10^{-6}(i,j=1,2)$  as shown in 
  eq.(\ref{feqcd}), eq.(\ref{feqcu}), eq.(\ref{feqhd}), and eq.(\ref{feqhu})
 needed for successful predictions of baryon asymmetry in this model.
   We provide below  how this difficulty can be circumvented in  two
   different ways :(i)Non-renormalizable, and (ii) Renormalizable; any
   one of these can be added to ${\cal L}^{(10)}$ for charged
   fermion mass parametrization.\\

\par\noindent{\bf (i). Non-Renormalizable Yukawa Correction:}\\
There have been attempts to represent fermion masses in SUSY SO(10) via
non-renermalizable interactions
with additional flavor symmetries and flavon fields  \cite{Dermisek:2006}.
Without introducing any such additional fields or symmetries, our
attempt here is confined to the non-SUSY SO(10) gauge symmetry and the Higgs
representations of the model. We note that the following non-renormalizable Yukawa (NRY) interactions are allowed\\
\bea    
{\cal L}^{(1)}_{NR}&=&\frac{F_{(1)}^{ij}}{M_{G}}{16}_i{16}_j {10}_H {45}_H, \nonumber\\
{\cal L}^{(2)}_{NR}&=&\frac{F_{(2)}^{ij}}{M_{G}^2}{16}_i{16}_j {10}_H {45}_H {45}_H.
\label{NRF}
\eea
where $M_G=$ Planck scale $M_{Planck}$, or the String scale $M_{String}$. The first Yukawa contribution
is suppressed by a factor ${M_{GUT}\over M_G}\sim 10^{-2}-10^{-3}$. Noting that ${10}_H\times
{45}_H \supset {120}_H\supset \xi(2,2,15)$, it contributes to
non-diagonal elements of all Dirac type mass matrices
antisymmetrically which we ignore in this qualitative explanation, but
can be included if a $\chi^2$ fit is desired in future works. The second Yukawa interaction
in eq.(\ref{NRF}) containing
${10}_H\times {45}_H\times {45}_H$ has an effective ${(2,2,15)}_H$
component that is contained in ${\bar {126}}$ and its contribution is symmetric. It is important to note that at the GUT scale ${\cal
  L}^{(2)}_{NR}$ gives a  suppressed  factor that adequately qualifies
it to parameterize  the needed additional corrections 
 with $m^0_{ij}\sim F_{(2)}^{ij}\frac{M^2_{GUT}}{M^2_G}v_{ew} \sim
F_{(2)}^{ij} (10^{-4}-10^{-5})v_{ew}$. Thus, at the GUT scale the quark
and lepton mass matrices can be  parameterized as:
\bea
M_u= G_u+F_u,\,\,  M_D=G_u-3F_u\,\,,\nonumber\\  
M_d=G_d+F_d,\,\,   M_l=G_d-3F_d\,\,,\label{GUTmass}
\eea  
where $G_u=Y^{(u)}<10_{H_u}>$, $G_d=Y^{(d)}<10_{H_d}>$,  $F_p \sim
F_{(2)}{10^{-4}}.<10_{H_p}>$ , $p=u,d$.
Details of fermion mass parametrization goes in a manner similar to
those discussed in \cite{PSB-rnm:2010,ap:2012,app:2013,bpn-mkp:2013,pas:2014}.

\par\noindent{\bf (ii). Renormalizable Correction:} 

Through renormalizable interaction, the improvement of fermion mass
parametrization  is also suggested by the
introduction of a second ${126}_H$ representation \cite{app:2013,pas:2014}. We denote this and
its corresponding components under $G_{224}$ as ${126}_H'\supset \Delta_L'(3,1, 10),
~\Delta_R'(1,3,{\bar {10}}),~\xi'(2,2,15),....$. In contrast to  the  $\Delta_L\subset {126}_H$ whose mass has
been fine tuned to be at $M_{\Delta_L}\sim 10^{12}$ GeV
for the implementation of the type-II seesaw component of the hybrid seesaw
 formula, leptogenesis, and coupling unification,
all the components of ${126}_H'$ are naturally assigned masses near
the GUT scale consistent with extended survival hypothesis
\cite{del Aguilla:1981,rnm-gs:1983}. Also no VEV is needed to be assigned
to  $\Delta_R'$ either i,e we fix $<\Delta_R'>=0$, since the
corresponding role of gauge symmetry breaking has been taken over by
$<\Delta_R(1,3, {\bar {10}})>=v_R\sim M_{GUT}\subset {\bar {126}}_H$. Thus the presence of the second Higgs representation ${126}_H'$ does not affect the type-II
seesaw and the RH neutrino masse parameters of type-I in the hybrid
seesaw formula of eq.(\ref{eq:nu_mass}). Even upto the two-loop level it
does not affect the gauge coupling unification of the present model. Denoting the corresponding SO(10)
invariant Yukawa term as $f'16.16. ({\bar {126}})'$, we have
renormalizable corrections to eq.(\ref{GUTmass}) where $F_u\to
F'_u=f'<\xi'_u>, F_d \to F'_d=f'<\xi'_d>$. It is well known that such
corrections provide reasonable parameterization of the fermion masses of
the first and second generations. With degeneracy of all superheavy
components of ${126}_H'$, its threshold corrections to unification
scale and proton lifetime are vanishingly small
\cite{RNM:1992}. Similarly, if the renormalizable antisymmetric
contributions to fermion mass matrices due to Yukawa interaction of a
${120}_H\subset SO(10)$ are included, its threshold effects on
unification scale and proton lifetime would be also vanishingly small
due to degeneracy of the components.

Alternatively the fermion mass
parametrization may be improved further by including both the
renormalizable and non-renormalizable contributions in
eq. (\ref{GUTmass}). In addition, the antisymmetric contribution
through the first nonrenormalizable term in ${\cal L}^{(1)}_{NR}$ may be
also included for still  further improvement. Further, the
antisymmetric NRY due to ${\cal L}^{(1)}_{NR}$ can be very well
replaced by renormalizable Yukawa contribution $h^{(120)}16.16.{120}_H$. 
 
 The next question is whether this parametrization significantly
 affects the predicted results of this work where we have used the
 boundary condition $M_D(M_{GUT})=M_u(M_{GUT})$. In SO(10) there are
 two maximal subgroups of rank $5$: the Pati-Salam group $G_{224}$ and
 the flipped $SU(5)\times {\tilde U}(1)(\equiv G_{fl})$. When $SU(4)_C\subset
 G_{224}$  is unbroken, the assumed boundary condition is
 exact. Similarly it is well known that in the presence of $G_{fl}$
 symmetry $M_D(M_{GUT})=M_u(M_{GUT})$. But in the process of SO(10) breaking to the SM, both these
 gauge symmetries are also broken  and the boundary condition is approximate
 to the extent that $M_u-M_D=4F_u$. This suggests that $\sigma_u\equiv
 4F_u/m_{\rm top}$ should be a small number in case fermion mass fit is also
 included as a required ingredient in this model. For a very preliminary
 estimation of $\sigma_u$ , we note the interesting point that the GJ
 relation $m^0_{\mu}=3m^0_s$ is almost exactly satisfied near the GUT
 scale $\sim 10^{15.56}$ GeV by values obtained in the bottom-up approach
 within the SM paradigm \cite{dp:2001}:
\bea
  m^0_{\mu}&\sim &93.14 \pm 0.01 \,MeV,\nonumber\\ 
m^0_s &\sim & 34.59 \pm 5.0 \, MeV. \label{GUTmdms}
\eea
With the dominance of the element ${(F_d)}_{22}$ in the $(22)$ elements of down-quark and charged lepton mass matrices,  
$|{(F_d)}_{22}|>>|{(G_d)}_{22}|$,  gives ${(F_d)}_{22} \sim 30$
MeV and a fractional change $\frac{(\Delta
  {M_D})_{22}}{({M^{u}}_D)_{22}}\sim 0.3$ compared to the uncorrected value
  of ${(M^{u}_D)}_{22}=262$ MeV shown in  Sec.2. 
We have checked that even afte applying these corrections satisfying
the first of GJ relation in eq.(\ref{GJrel}), our solutions and
predictions on baryon asymmetry made in this work are not
significantly affected. Also they  remain largely
unaffected as long as the corrections to the elements of the Dirac neutrino mass
matrix $M_D$ are either less or at
most of the same order as those given in Sec.2. After the GUT symmetry
breaking to the SM gauge theory we have assumed only one linear combination of
different up type and down type doublets to remain massless to form
the standard Higgs doublet.

\section{Acknowledgment}
M. K. P. acknowledges financial support through the research project
SB/S2/HEP-011/2013 awarded by the Department of Science and Technology,
Government of India. R. L. A. acknowledges the award of a
Post-Doctoral Fellowship by Siksha 'O' Ausandhan University
where this work was carried out.

\end{document}